\documentclass[notitlepage,superscriptaddress,amsmath,amssymb,aps,pra,twocolumn]{revtex4-2}
\usepackage{amsmath,amssymb,graphicx,mathtools,dsfont,ulem,bbold,hyperref,soul}
\usepackage[dvipsnames]{xcolor}
\usepackage[T1]{fontenc}
\hypersetup{citecolor=red,colorlinks=true,urlcolor=blue}

\usepackage{ulem}
\usepackage{soul}
\usepackage{comment}

\newcommand{\br}    { \mathbf{r}}
\newcommand{\ketbra}[2]{|#1\rangle\!\langle#2|}
\newcommand{\av}[1]    {\langle #1 \rangle}
\newcommand{\modsq}[1]    {\left| #1 \right|^2}
\newcommand{\ket}[1]{\mbox{$| #1 \rangle$}}
\renewcommand{\Tilde}[1]{\hat{\tilde#1}}
\newcommand{\NCU}{Institute of Physics, Faculty of Physics, Astronomy and Informatics, Nicolaus Copernicus University in Toru\'n, Grudzi\c{a}dzka 5, 87-100 Toru\'n, Poland}
\newcommand{\FUW}{Faculty of Physics, University of Warsaw, ul. Pasteura 5, PL–02–093 Warszawa, Poland}
\newcommand{\braket}[2]{\langle #1 | #2\rangle}
\newcommand{\tr}[1]    {{\rm Tr}\left[ #1 \right]}
\newcommand{\bra}[1]{\mbox{$\langle #1 |$}}
\newcommand{\re}[1]    {\mathrm{Re}\left[#1\right]}

\usepackage{soul,ulem}
\newcommand{\twm}[1]{#1}

\begin{document}

%\title{\twm{Dynamical} generation of many-body entanglement\\ by collective coupling of atom pairs to cavity photons
%\twm{Engineering interactions by collectively coupling\\atom pairs to cavity photons for quantum metrology}
%\twm{Engineering Photon-Induced Atom-Atom Interactions\\ via Cavity-Coupled Atom Pairs for Efficient Many-Body Entanglement}

%\title{Cavity-photon induced atom-pair interactions for \\ quantum metrology and Bell nonlocal correlations}

\title{{Engineering interactions by collective coupling of\\atom pairs to cavity photons for entanglement generation}}

\author{Sankalp Sharma}\affiliation{\NCU}
\author{Jan Chwedeńczuk} \affiliation{\FUW}
\author{Tomasz Wasak}\affiliation{\NCU}

\begin{abstract}
% New version: more emphasis on interaction engineering
\twm{Engineering atom-atom interactions is essential both for controlling novel phases of matter and for efficient preparation of many-body entangled states, which are key resources in quantum communication, computation, and metrology. In this work, we propose a scheme to tailor these interactions by coupling driven atom pairs to optical cavity photons via a molecular state in the dispersive regime, resulting in an effective photon-field-dependent potential. As an illustrative example, by analyzing the quantum Fisher information, we show that such induced interactions can generate robust many-body entanglement in two-mode ultracold bosons in an optical cavity. By tuning the photon-induced interactions through the cavity drive, we identify conditions for preparing highly entangled states on timescales that mitigate decoherence due to photon loss. Our results show that entanglement formation rate scales strongly with both photon and atom number, dramatically reducing the timescale compared to bare atomic interactions. We also identify an optimal measurement for exploiting the metrological potential of the atomic state in an interferometric protocol with significant photon losses, saturating the quantum Cramer-Rao lower bound. Furthermore, we show that despite these losses the atomic state exhibits strong Bell correlations. Our results pave the way for engineering atom-atom interactions to study novel phases of light and matter in hybrid atom-photon systems, as well as for tailoring complex quantum states for new quantum technology protocols and fundamental tests of quantum mechanics.}
\end{abstract}

\maketitle

{\it Introduction} --- \twm{Scalable many-body entangled states are essential resources for quantum technologies of the future, such as quantum communication, computation and metrology.  The generation of these states relies on precisely controlled interactions between the particles. They should enable fast creation of quantum correlations in order to mitigate the effects of decoherence, which is a fundamental obstacle even in most advanced equipments operating in well-isolated environments. In degenerate quantum gases, for instance, the formation of entanglement can be significantly accelerated by exploiting Feshbach resonances which increase the rate of two-body collisions~\cite{chin2010feshbach, inouye1998observation, timmermans1999feshbach}. This enhancement, however, comes at a cost --- the magnetic field fluctuates, and, as the collisions intensify, the three-body losses come into play~\cite{massignan2008efimov, ratzel2021decay, stenger1999strongly}.}

\twm{A promising approach for generating entangled states with tailored properties is based on cavity quantum electrodynamics (QED) with ultracold atoms.  This platform is particularly advantageous because it enables independent control over the light and matter components of the hybrid system. Such control allows for a strong coupling regime of the atom-photon interaction and the engineering of entangled states with specific desired properties}~\cite{haroche2006exploring}. 
Moreover, cavity QED systems can be probed in real time via transmission spectra~\cite{mekhov2007probing}. They are also a basis for the development of new non-equilibrium dissipation-controlled quantum dynamics protocols~\cite{dogra2019dissipation}. Finally, they enable quantum simulations of solid-state Hamiltonians and non-equilibrium effects in complex systems beyond the schemes known from condensed matter~\cite{mivehvar2021cavity}. 
\twm{These unique possibilities critically depend on tunable-range photon-mediated atom-atom interactions}~\cite{munstermann2000observation,vaidya2018tunable,norcia2018cavity,mottl2012roton, mivehvar2023tuning}. 
The quantum dynamics of self-organization of atoms in an optical cavity was shown to create strong atom-photon entanglement that is, however, fragile to photon losses~\cite{maschler2007entanglement, vukics2007microscopic}.
Recently, such photon-induced interactions in a cavity have been used experimentally to generate spin- and momentum-correlated atom pairs in a Bose gas~\cite{esslinger2024spin}. Also, photon-atom scattering in a pumped ring cavity, causing the transverse self-organization of bosons, led to momentum multiparticle entangled Dicke-squeezed states~\cite{ackemann2023generating}.

Despite the versatility of the effects induced by the cavity, such as the generation of novel phases of matter~\cite{baumann2010dicke,ritsch2013cold,klinder2015observation, landig2016quantum, leonard2017supersolid, leonard2017monitoring}, the photon-matter interaction has so far been limited to the dipole coupling of atoms, in which a single photon interacts with a single atom from a many-body system. However, a recent experiment reported the observation of universal pair-polaritons in a strongly interacting two-component Fermi gas~\cite{konishi2021universal}. Crucially, in this setup, the photons from a high-finesse optical cavity were directly coupled to pairs of atoms via a molecular state, which is in contrast to the regular scheme of dipole coupling where a single atom is excited by a single photon.
This mechanism enabled a real-time weakly destructive probe of the pair correlation function of the atoms, and it holds promise for the development of novel methods to engineer atomic interaction potentials. 

\twm{Here, we exploit this light-matter coupling to derive the form of the photon-induced atom-atom interactions under different system driving schemes. Based on this approach, we propose a novel method for the fast generation of many-body entangled states on a time scale controlled by the strength of the drive. The setup consists of  ultracold bosons trapped in a double-well potential immersed in an optical cavity. Our results show that photon-induced atom-atom interactions lead to the generation of highly entangled atomic states which may trigger the development of new protocols and applications in quantum technologies}~\cite{mivehvar2021cavity, gietka2015interferometry, niezgoda2021cooperatively, ackemann2023generating,esslinger2024spin}.

\begin{figure}[t!]
  \centering
  \includegraphics[width=\linewidth]{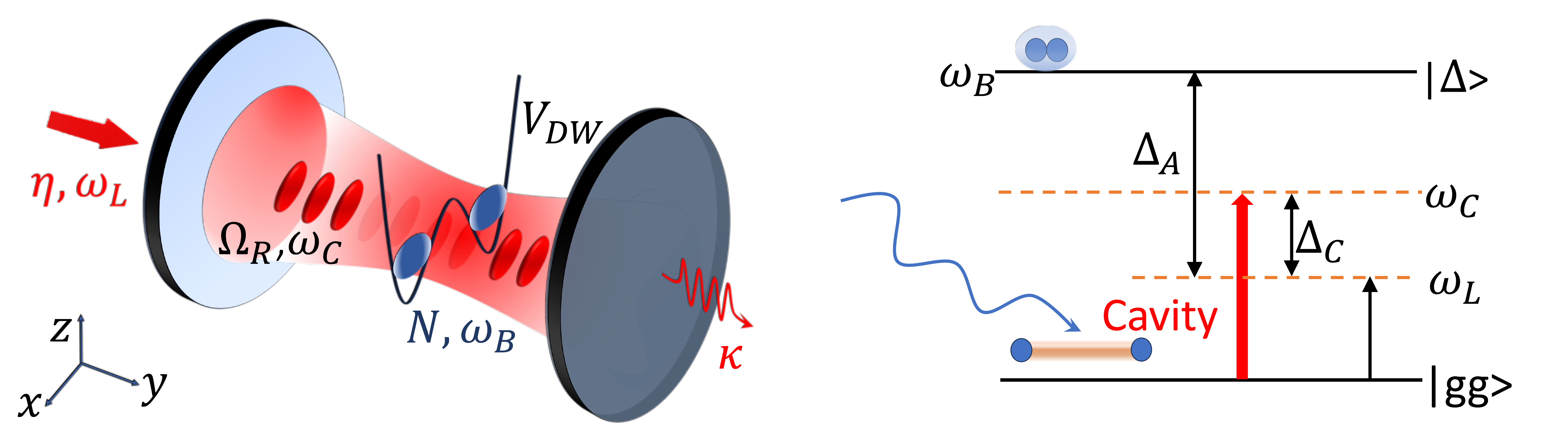}
    \caption{
    {\bf Left}: Illustration of the setup. $N$ \twm{ultracold atoms with bosonic statistics} are confined in a double-well potential $V_{DW}$ inside an optical cavity. 
    The laser pump $\eta$ with frequency $\omega_L$ enters from one of the mirrors of the cavity with photon loss rate $\kappa$. 
    The coupling between the atomic cloud and the cavity field is captured by the single-mode {coupling strength} $\Omega_{R}$. 
    {\bf Right}: The diagram of the relevant energy levels of atomic pairs $\ket{gg}$ and the molecule $\ket{\Delta}$.
    By $\Delta_A$ and $\Delta_C$ we denote detunings from the molecular state with energy $\omega_B$ and cavity photon energy $\omega_c$, respectively.
  }\label{fig:scheme}
\end{figure}

\twm{{\it Photon---atom--pair interaction in a cavity}}---
Our system of interest is a collection of $N$ atoms of mass $m$ \twm{held in a trap $V(x)$ in an optical cavity (see Fig.~\ref{fig:scheme} with an example of a double-well potential)}. 
The atoms are in the ground state interacting through a contact potential of strength $g_{0}=4\pi a/m$, where $a$ is the $s$-wave scattering length (we use $\hbar=1$ units). 
We assume that the pairs of atoms have an optically accessible molecular state with energy $\omega_{B}$~\cite{konishi2021universal}.

The total Hamiltonian is $\hat H = \hat H_A\!+\! \hat H_\Delta\!+\!\hat H_C \!+\! \hat H_{CA} \!+\! \twm{\hat H_P}$, where the terms describe ground state atoms, molecules, cavity photons, light-matter interactions, \twm{and external pump}, respectively~\cite{supp}.
The atoms are governed by
\begin{align}\label{HA}
  \hat{H}_{A}\! \!=\!\!\int d^3r\, \hat{\Psi}^\dagger\left[-\frac{\nabla^2}{2m}\!+\!V\right]\hat{\Psi}
 + \frac{g_0}{2}  \int \!\!d^3r\,\hat{\Psi}^{\dagger}\hat{\Psi}^\dagger\hat{\Psi}\hat{\Psi},
\end{align}
%\begin{align}\label{2}
%  \hat{H}_{A} &=\int d^3r\, \hat{\Psi}^\dagger_{g}(\br)\left[-\frac{\nabla^2}{2m}+V(\mathbf{r})\right]\hat{\Psi}_{g}(\br)\nonumber \\ 
%  & + \frac{g_0}{2}  \int \!\!d^3r\,\hat{\Psi}^\dagger_{g}(\br)\hat{\Psi}^\dagger_{g}(\br)\hat{\Psi}_{g}(\br)\hat{\Psi}_{g}(\br),
%\end{align}
\twm{where $\hat\Psi(\br)$ is the ground-state atom field operator for bosons~\cite{mivehvar2021cavity, mivehvar2023tuning}, i.e, its commutator satisfies $[\hat\Psi(\br), \hat\Psi(\br')] = \delta(\br-\br')$, with $\delta(\br)$ the Dirac delta}, and $V(\br)$ is the trapping potential.
For the molecular state \twm{of the pairs of atoms}, we assume that they are tightly confined and effectively described by
\begin{align}\label{3}
  \hat{H}_\Delta = \int d^3 r \hat\Psi_\Delta^\dagger(\br) \left[ -\frac{\nabla^2}{2 M} + V(\mathbf{r}) + \omega_B \right] \hat\Psi_\Delta(\br),
\end{align}
where $M=2m $ is the molecule mass. 

\twm{The high-$Q$ optical cavity with the axis along $y$-direction is characterised by the resonant frequency $\omega_{c}$ and the mode of light is governed by the Hamiltonian $\hat{H}_C =  \omega_c \hat{a}^{\dagger} \hat{a}$, where $\hat a$ is the annihilation operator of the cavity photon with a position-dependent mode function $f(\br)$~\cite{supp}. We model the atom-light interaction by the Hamiltonian in the rotating wave approximation,
\begin{align}\label{HCA}
  \hat{H}_{CA} \!=\! -i\Omega_{R}\!\! \int\!\! d^3r f(\br)\!\!\left[\hat\Psi_\Delta^\dagger(\br) \hat\Psi^2(\br) \hat{a}  -  \mathrm{H.c.}\right],
\end{align}
where $\Omega_R$ sets the single--photon---single--pair coupling strength, see also SM~\cite{supp}, while $\mathrm{H.c.}$ is the Hermitian conjugate. $\hat H_{CA}$ describes a process (and its reverse), see Fig.~\ref{fig:scheme}, where a single photon annihilates a pair of ground state atoms leading to a formation of a single molecule.} Such a pair-molecule coupling mechanism~\cite{holland2001resonance}, although without light quantization, was discussed in Ref.~\cite{timmermans2001prospect} in the context of composite Fermi–Bose superfluids and pairing by pair exchange through a molecular condensate. For bosons, it has been used to study the quantum dynamics of the photoassociation of molecules in an optical cavity from a single-mode Bose-Einstein condensante~\cite{zivkovic2009photoassociation}. \twm{Recently, Ref.~\cite{konishi2021universal} exploited this coupling to observe universal pair-polariton formation (i.e., independent of the specific molecular transition used) in a strongly interacting two-component Fermi gas.}

\twm{Below, we assume the cavity is pumped by a laser with a frequency $\omega_{L}$, detuned from the cavity frequency by $\Delta_c = \omega_L - \omega_C$, through one of its mirrors. Additionally, in analogy with the regular single--photon---single--atom case of the dipole coupling~\cite{mivehvar2021cavity},  the direct transversal pumping of the molecular transition can be employed.  Thus, the pumped single mode of radiation inside the cavity is described by 
$\hat H_P = -i  \eta e^{-i \omega_L t} \hat a^{\dagger} - i \Omega_R \int d^3r\,\hat\Psi_\Delta^\dagger \hat \Psi^2 \eta_\mathrm{t} + \mathrm{H.c.}$, where $\eta$ denotes the driving strength 
  through the mirrors and $\eta_\mathrm{t}(\br)$ is the strength of the transversal pump mode.}

\twm{
If the atomic detuning $\Delta_A = \omega_L-\omega_B$ is much larger than the other characteristic frequency scales of the system,  the population of the molecular state is small and follows the ground state adiabatically, and thus can be eliminated. Then the effective Hamiltonian of the system is given by $\hat H$, where $\hat H_\Delta + \hat H_{CA}$ is replaced by the photon-induced interaction
\begin{align}\label{HCAA}
%  \hat{H}_{AA}^C \!=\! U_0 \int \!\! d^3r\, \hat a^\dag \hat a [f(\br)]^2 
%  \hat\Psi^\dag(\br)\hat\Psi^\dag(\br)\hat\Psi(\br)\hat\Psi(\br),
    \hat{H}_{AA}^C \!=\! U_0 \int \!\! d^3r\, [f^*(\br)\hat a^\dag + \eta_\mathrm{t}][f(\br)\hat a + \eta_\mathrm{t}] 
    \hat\Psi^\dag\hat\Psi^{\dag}\hat\Psi\hat\Psi,
\end{align}
where $U_0 = \Omega_R^2/\Delta_A$ is the strength of the atom-atom interaction per photon induced by cavity photons~\cite{supp}. 
Unlike conventional single-photon--single-atom coupling, where cavity photons mediate long-range atomic interaction after adiabatic elimination in the dispersive regime~\cite{munstermann2000observation,vaidya2018tunable,norcia2018cavity,mottl2012roton, mivehvar2021cavity}, the interaction $\hat{H}_{AA}^C$ retains the operator and dynamical character of cavity photons and local position dependence via the mode function. This enables the generation of highly entangled states and complex quantum dynamics offering exciting opportunities for quantum technology applications~\cite{degen2017quantum}.
}

\twm{{\it Cavity-assisted bosonic Josephson junction}}---
\twm{To illustrate how the interaction from Eq.~\eqref{HCAA} influences the buildup of atomic correlations, we consider a bosonic Josephson junction (BJJ) with $N$ atoms, without transverse driving, i.e., $\eta_\mathrm{t}=0$. We expand the field operator $\hat{\Psi}(\br)$ into two modes, $\hat b_1$ and $\hat b_2$, with associated spatial mode functions localized around the minima of the double-well potential, see Fig.~\ref{fig:scheme}. In the rotating frame of the laser, we derive an effective atom-photon Hamiltonian $\hat H_\mathrm{eff}= -\tilde\Delta_c \hat a^\dag \hat a - J \hat J_x + U \hat J_z^2 + 2W_0 \hat a^\dagger \hat a \hat J_z^2$, where the last term originates from Eq.~\eqref{HCAA} and the detuning is renormalized as $\tilde\Delta_c = \Delta_c - W_0 N^2/2$ \cite{supp}; by $\hat J_i$, where $i=x,y,z$, we denote the Schwinger-boson representation of the modes $\hat b_{1/2}$~\cite{gross2012spin, supp}. Here, $J$ is the tunneling amplitude, and $U$ is the on-site energy while $W_0$ determines the strength of the photon-induced atomic interactions~\cite{supp}. Importantly, the cavity modifies the atom-atom interaction in contrast to the reported cavity-assisted tunneling in the regular single-photon--single-atom coupling~\cite{szirmai2015tunneling}.}

{\it Generation of many-body entanglement \twm{in BJJ}}---\twm{We now demonstrate that the effective Hamiltonian, $\hat H_\mathrm{eff}$, generates strong entanglement between atoms. To illustrate this, we consider a scenario analogous to the one-axis twisting (OAT) protocol}, which has been successfully employed to create many-body entangled states in two-mode atomic configurations~\cite{esteve2008squeezing,gross2010nonlinear,riedel2010atom,berrada2013integrated}. \twm{Here, all atoms are initially placed in a single mode, 
and then the bare Josephson oscillation (governed by tunneling and in the absence of other terms) creates a separable state in which all
atoms are in coherent spin state~\cite{gross2012spin} with a symmetric superposition of the two modes, namely,
$\ket{\psi_0}\propto(\hat b_1^\dagger+\hat b_2^\dagger)^N \ket{\mathrm{vac}}$.}

\twm{Next}, the Josephson term is suppressed by raising the inter-well barrier (and thus setting $J=0$) and the atomic interactions correlate the particles. 
Without the cavity, this procedure is the standard OAT, because the Hamiltonian $\propto \hat J_z^2$ reduces atomic fluctuations, creating spin-squeezing~\cite{wineland1994squeezed,sorensen2001many,
gietka2015interferometry}.

\twm{The standard OAT process is governed by two time scales. 
First, after $t\propto N^{-2/3}$, the Gaussian spin-squeezed states are created which yield sub-shot-noise sensitivity~\cite{supp}, i.e.,  $\Delta\theta < 1/\sqrt N$, of phase estimation in the Mach-Zehnder interferometer (MZI)~\cite{esteve2008squeezing,pezze_mzi}.  Later, when $t\propto N^{-1/2}$ the system enters a non-Gaussian  regime~\cite{pezze2009entanglement,smerzi_ob,pezze2018quantum}, where 
the many-body entangled states have to be characterized by the quantum Fisher information (QFI)~\cite{braunstein1994statistical}, i.e.,}
\begin{align}\label{eq.qcrb}
  F_Q = 2 \sum_{i,j}\frac{\left(\lambda_{i}- \lambda_{j}\right)^2}{\lambda_{i}+\lambda_{j}}\left|\langle \Psi_{i}|\hat J_y| \Psi_j\rangle\right|^2,
\end{align}
which sets
the Cramer-Rao lower bound (CRLB): $\Delta\theta\geqslant F_Q^{-1/2}$. Here,
 $\hat J_y$ \twm{is the generator of the MZI transformation~\cite{gross2012spin,esteve2008squeezing}, and the sum runs over the spectrum of the atomic density matrix, i.e., $\hat\varrho_A = \sum_i \lambda_i \ketbra{\Psi_i}{\Psi_i}$. In the non-Gaussian regime, very strongly entangled states manifest through the Heisenberg scaling $F_Q\propto N^2$~\cite{hyllus2012fisher}.}

\twm{We now demonstrate that even in the presence of photon losses, the light-induced atom-atom interaction from Eq.~\eqref{HCAA} 
  can drastically accelerate the generation of strong entanglement compared to the above-mentioned rates characterising the OAT procedure.}
\twm{Initially, atoms are in a state $\ket{\psi_0}$ and the photonic vacuum is dynamically populated by $\hat H_P$ with $\eta>0$.
In the Supplementary Materials we discuss the other case, when photons are initially in a coherent state and pumping is absent~\cite{supp}. This scenario 
is similar to the usual optical Feshbach resonance in free space with a coherent beam~\cite{chin2010feshbach}, that  can lead to dynamical instabilities of atoms~\cite{wasak@2013nonlinear}.} 

\begin{figure}[t!]
  \centering
        \includegraphics[width=1.0\linewidth]{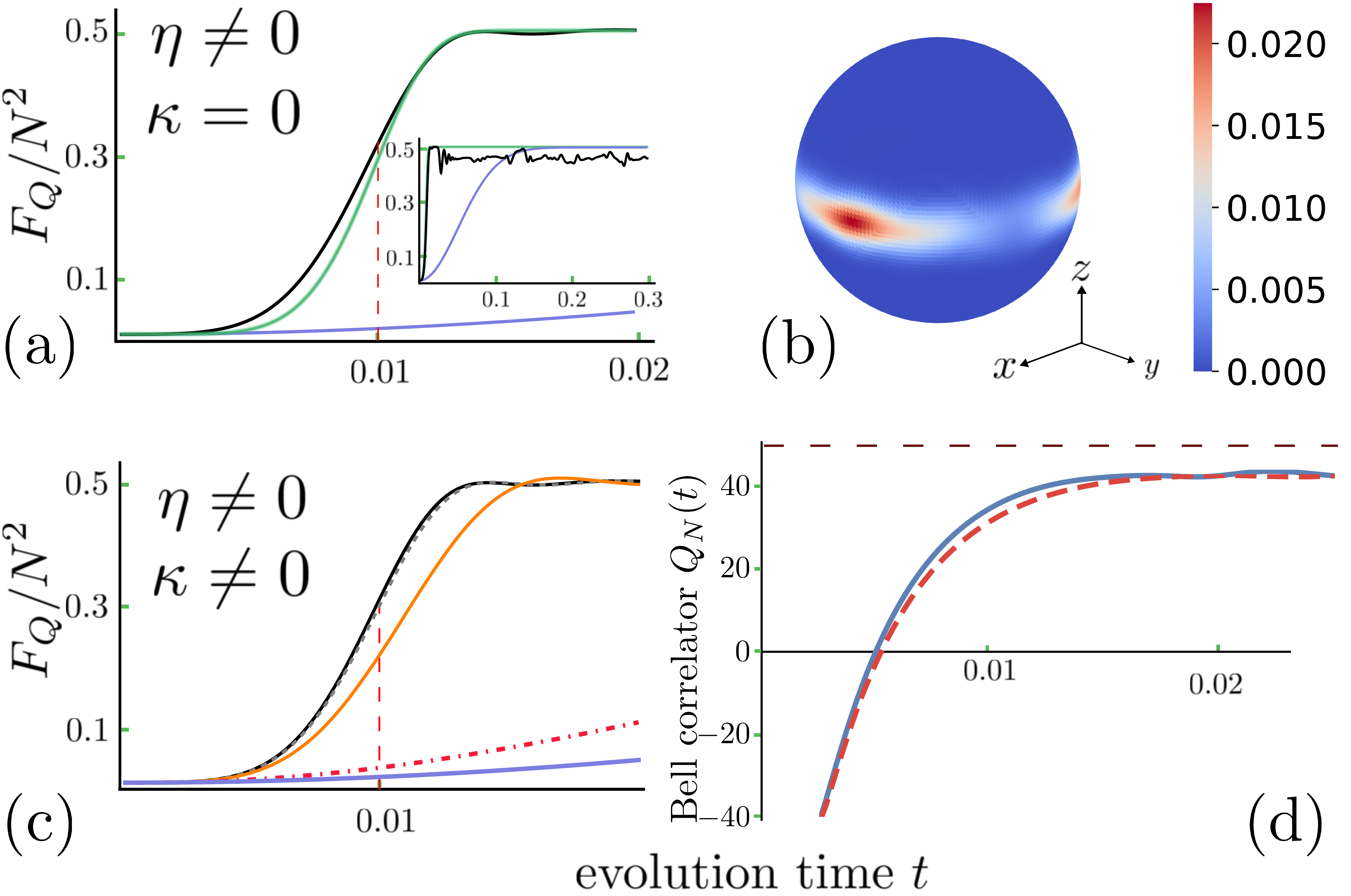}
        \caption{
        %
        % TODO:
        %\twm{\textbf{[---TODO}: labels: upper left=(a), upper right=(b), lower left=(c), lower right=(d); color code for Husimi; unit of time; label for the vertical axis for Bell\textbf{---]}\\}
        %
        %
        \twm{
        \textbf{(a, c)} QFI as a function of time (in units of $U^{-1}$), without (a) and with (c) photon losses. The black line shows $F_Q$ for $\kappa=0$, and the blue line for the OAT method. Inset shows long-time behavior. In (a), the solid green line is QFI from Eq.~\eqref{eq.qfi.approx}. In (c), QFI is for $\kappa \tau_Q = 0.1$ (dashed grey), 1 (orange), and 10 (dot-dashed red). The vertical dashed red line indicates $\tau_Q=0.01 U^{-1}$.
        \textbf{(b)} Husimi distribution at $t = 1/(4U\sqrt{N})$ shows a highly non-Gaussian state of atoms, $\kappa=0$. 
        Parameters for all panels: $U=W_0=\Delta_c=1$, $N=100$ atoms, pump $\eta=320$.
        \textbf{(d)} Bell correlator for $N=50$ in the lossless case (blue solid line) and with losses  $\kappa \tau_Q = 1$ (red dahsed line). The black horizontal dashed line denotes the maximal possible value of $Q_N(t)$.
        All values of $Q_N(t)>0$ indicate the presence of Bell correlations.
        }
        }
    \label{fig.total}
\end{figure}

%The density matrix of the atomic degrees of freedom at time $t$~\cite{supp} is
\twm{For the lossless case,}
an analytical expression for the atomic density matrix at time $t$, after tracing-out the photonic degrees of freedom, reads
\begin{align}\label{eq.dens.gen}
    \hat\varrho_A(t)
    \!\!=\!\!\!\sum_{m,m'}C_mC_{m'}e^{-iU(m^2-m'^{2})t}u_{mm'}(t)\ketbra{m}{m'},
\end{align}
where the time-dependent function captures the effect of photon-induced interactions:
\begin{align}\label{eq.dens.at}
    u_{mm'}(t)&=e^{-\frac12(\modsq{\gamma_m}+\modsq{\gamma_{m'}})}e^{\gamma_m\gamma^*_{m'}}e^{-i\eta t(\beta_m-\beta_{m'})}\times\nonumber\\
    &\times e^{-i\beta^2_m\sin(\omega_mt)}e^{i\beta^2_{m'}\sin(\omega_{m'}t)}%\label{eq.red.at},
\end{align}
with $\omega_m=-\Delta_c+2W_0m^2$, $\beta_m=\eta/\omega_m$ and $\gamma_m=\beta_m(e^{-i\omega_m t}-1)$, see SM for a detailed derivation~\cite{supp}). This result is subsequently employed to quantify the entanglement strength between the atoms.

\twm{Although the calculation of the QFI  for mixed states is typically challenging or even intractable, we found an analytical expression in the large-$N$ limit under the following assumptions~\cite{supp}.}
\twm{Namely, if the bare atom interactions can be neglected, i.e., $W_0\bar n(t)\gg U$, where $\bar n(t)$ is the average number of photons, then in the short-time regime, defined by $\sqrt{N}W_0 t \ll 1$, where atomic coherences from the interaction remain minimal, the condition $\bar n \gg 1$ yields the following QFI}:

%in the short-time regime, $\sqrt{N}W_0 t \ll 1$, when the atomic coherences induced by the interaction are small, the $\bar n\gg1$ limit yields the QFI:}

%\sks{Maybe we can rephrase the above like this: 

%Namely, if the interactions between bare atoms can be disregarded—meaning that $W_0\bar n(t) \gg U$, where $\bar n(t)$ represents the average photon number—then in the short-time regime, defined by $\sqrt{N}W_0 t \ll 1$, where atomic coherences from the interaction remain minimal, the condition $\bar n \gg 1$ yields the following QFI:}
%
%\twm{Specifically, the time $t$ must be short enough for $4W_0t\sqrt N\ll1$ to hold. Then, the atom-light coupling must dominate over the purely atomic interaction, i.e., $W_0\bar n(t)\gg U$, where $\bar n(t)$ is the average number of photons. Finally, the average number of photons must be large, namely $\bar n\gg1$.} 
%
%If these conditions are satisfied, the QFI for the atomic subsystem is given by~\cite{supp}
\twm{
\begin{align}\label{eq.qfi.approx}
  F_{Q} \simeq \frac{N}{2}[1 + f_Q(t)] + \frac{N^2}{2}[1-f_Q(t)],
\end{align}
}
with the dynamical crossover function reads
  \begin{align}
    &\twm{f_Q(t)} = \exp\left[-2 \left( \frac{t}{\tau_Q}\right)^6\right],\ \ 
    \twm{\tau_Q}
    = \left(\frac9{4N W^2_0\eta^4}\right)^{\frac{1}{6}}. \label{f_b}
  \end{align}
%\twm{Note that the average number of photons changes for short times as $\bar n(t)=\eta^2t^2$, and, therefore, $f_Q$ can be expressed as a Gaussian function of time .}
The value of the QFI grows rapidly from the shot-noise limit, i.e., $F_Q=N$, at $t=0$ to half of Heisenberg limit $F_Q=N^2/2$ on a time given by $\tau_Q$. 
%\twm{\st{Compared to OAT, the scaling is weaker and strong entanglement is generated on a scale $\tau_Q\propto N^{-1/6}$. In SM we show that, for initial coherent state of light, the timescale $\tau_Q \propto N^{-1/2}$ similar to OAT.}}
\twm{Crucially, the timescale $\tau_Q$ can be shortened by injecting photons into the cavity. Additionally, the effective coupling strength that determines $\tau_Q$ is proportional to $ \sqrt{N}W_0$, a result of the collective coupling between atoms and photons.}

In Fig.~\ref{fig.total}b, we display Husimi distribution~\cite{meystre2006coherence,schleich2011quantum,julia2012dynamic}, which is a positive quasi-probability distribution used to represent quantum states in phase space \cite{lee2015visualizing}. Specifically, for a two-mode atomic density matrix 
$\hat\varrho_A$ (see SM~\cite{supp}), the Husimi $Q$ function is $Q(\theta, \phi) = \langle \theta, \phi | \hat\varrho_A | \theta, \phi \rangle$,
where $|\theta, \phi \rangle$ is a spin-coherent state expressed in terms of the bosonic creation operators as: $|\theta, \phi \rangle = \frac{1}{\sqrt{N!}} \left( \cos\frac{\theta}{2} \, \hat{b}_1^\dagger  + \sin\frac{\theta}{2} \, e^{i \phi}\hat{b}_2^\dagger  \right)^N |0 \rangle$ ~\cite{johansson2012qutip}.

%for the density matrix of the atomic subsystem. 
The rapid cavity dynamics result in a non-Gaussian state distributed around the equator of the Bloch sphere.
Fig.~\ref{fig.total}a shows the QFI from Eq.~\eqref{eq.qcrb} (black) compared with the approximate formula Eq.~\eqref{eq.qfi.approx} (green) and juxtaposed with the QFI for the OAT, i.e., $W_0=0$.
Here, we take $\eta=320 U$ so that at $t=0.02U^{-1}$, the average number of photons
is $\bar n(t)\approx40$, hence the pumping is fast enough to compete with the time-scale $\tau_Q$ of the entanglement build-up. We take $N=100$ atoms and the other parameters are set to unity, i.e., $W_0/U=\Delta_c/U=1$.
Although in a recent experiment~\cite{konishi2021universal}, $W_0$ was less than $U$, the observation of polaritons in this setup is an indication that $W_0 \sim U$ should be achievable~\cite{supp}.

The approximation works exceptionally well recovering the time-scale of entanglement build-up and the saturation level. 
\twm{In SM~\cite{supp}, we demonstrate similar agreement for the coherent state case.}
The QFI shows a flat build-up, as the number of photons needs to accumulate to start driving the entanglement growth, manifesting in a large exponent in Eq.~\eqref{f_b}. Nevertheless, for a moderate number of photons, the improvement over the bare OAT procedure is impressive. The inset shows the long-time behavior. While the approximation, treating $N$ as a continuous variable, cannot capture long-time oscillations (similar to the coherent-state approximation in the Jaynes-Cummings model~\cite{knight_qo}), it is sufficient to state that the photon coupling to atom pairs in the dispersive regime significantly enhances metrologically useful many-body entanglement.

{\it Impact of photon losses} --- \twm{We now investigate the impact of photon lossess on atomic entanglement. The QFI in Eq.~\eqref{eq.qfi.approx} is given by the average number of photons, suggesting that on the time scale considered, $F_Q$ is weakly sensitive to photonic decoherence.}

% To confirm this obsservation, This observation is confirmed below, demonstrating a major advantage of the considered setup in controlling the strength of entanglement.

\twm{We model the photon losses by the Lindblad master equation~\cite{manzano2020short} governing the evolution of the system:
\begin{align}\label{eq:M_E}
  \partial_t\hat\varrho=-i\left[\hat H_{\rm eff},\hat\varrho\right] +\kappa\left(\hat a\hat\varrho a^\dagger-\frac12\left\{\hat\varrho,\hat a^\dagger \hat a\right\}\right),
\end{align}
where $\kappa$ is the loss rate and $\{\,,\,\}$ stands for the anticommutator~\cite{johansson2012qutip}. 
For the same initial state as in the lossless case, we now use $\kappa \tau_Q = 0.1, 1, 10$, so that losses span three orders of magnitude of the scale $\tau_Q$ that sets the entanglement build-up time. Fig.~\ref{fig.total}c confirms that as long as $\kappa \tau_Q\lesssim1$, $F_Q$ remains roughly unaffected~\cite{supp}.}
For the strong loss $\kappa\tau_Q\gg 1$, the initial growth slows down --- the leakage of photons leaves the system in a bare OAT mode.
Decoherence in the photonic subspace does not hinder the build-up of atomic entanglement, as Heisenberg scaling of $F_Q\sim N^2$ is preserved even when $\kappa \tau_Q \sim 1$. However, it does influence the timescale over which entanglement is generated. Despite significant losses, the cavity-assisted OAT protocol remains effective in producing strong, genuine multi-particle entanglement. \twm{Moreover, after the entanglement is generated, the atomic state can be exploited for metrological purposes. Our calculations for $\kappa \tau_Q \sim 1$ show that $F_Q$ remains large over a long timescale after the pump is switched off at $t \sim 2 \tau_Q$, despite the rapid depletion of cavity photons~\cite{supp}.}

Finally, we note that recent observations of pair-polaritons in the strong coupling regime~\cite{konishi2021universal} for long observation times of the order of $0.5$ s with frequent interrogations of the system with a probe beam indicate that, even in the dispersive regime considered in this work, where only virtual molecular transitions are involved, the effect of three-body losses can be neglected. \twm{Also, since the photons are far off-resonance from the molecular transition, the molecular population is significantly suppressed minimizing the heating of the atomic gas due to spontaneous emission. We expect that photon loss from the cavity is the dominating source of noise, and we leave the impact of the heating for future research.}

\twm{{\it Optimal measurement} --- A physical measurement always yields a classical Fisher information $F\leqslant F_Q$, where $F = \sum_m p_m(\theta)(\ln\partial_\theta p_m(\theta))^2$, and the set of measurement operators $\{ \hat\Pi_m\}$ defines the probabilities $p_m(\theta) = \mathrm{Tr[\varrho_A^\theta\hat\Pi_m]}$~\cite{braunstein1994statistical, fluct_smerzi, smerzi_ob, cappellaro2017sensing}. 
%
%Husimi distribution in Fig.~\ref{fig.total}b indicates that rotation about $y$-axis by $\pi/2$ yields large peaks in the distribution of $\hat J_z$, indicating high sensitivity to rotations about $\hat J_y$.
We find that the entanglement of $\hat\varrho_A$ can be extracted by first rotating $\hat\varrho_A$ by $\hat V=e^{i \vartheta \hat J_x}$ and then measuring population imbalance of the wells, see SM~\cite{supp}, i.e., the measurement is $\hat\Pi_m = \hat V \ketbra{m}{m}\hat V^\dagger$ with $\hat J_z\ket{m} = m\ket{m}$.
Remarkably, selecting a single $\vartheta\approx \pi/2$ for all times, $F$ is indistinguishable from $F_Q$ in Fig.~\ref{fig.total}a, even with losses $\kappa\tau_Q\sim1$~\cite{supp}. The value of $F$ can be extracted by the Hellinger distance method developed in~\cite{strobel2014fisher}.
%\sks{I think we can also discuss here about the extraction of entanglement and relate it with CFI, moreover we can also cite the following reference \cite{strobel2014fisher}.}
%\twc{I added a sentence on the Hellinger distance method (the final one above).}
}

\twm{{\it Many-body Bell correlations}} \twm{--- Finally, we demonstrate that, in addition to strong entanglement useful for metrology, the atoms exhibit many-body Bell correlations~\cite{bell_local}. 
  To this end, we use a Bell inequality developed in Refs.~\cite{PhysRevLett.126.210506, JC2022, Plodzie2023generation}, namely $Q_N(t)\leqslant0$, where $Q_N(t) \equiv \ln[2^N \mathcal E_N(t)/(N!)^2]$, with $\mathcal E_N(t) \equiv |\mathrm{Tr}[\varrho_A(t) \hat J_+^N]|^2$ and $\hat J_+ = \hat J_z + i \hat J_y$.  Fig.~\ref{fig.total}d shows that after the time $\sim\tau_Q/2$, $Q_N(t)$ becomes positive 
{even in presence of losses}, signalling generation of Bell correlation. Moreover, for $t\gtrsim\tau_Q$, $Q_N(t) \!\sim\! N$ which indicates that $\sim N$ atoms become Bell--correlated. 
{$Q_N$ can be experimentally extracted by measuring multiple quantum coherences technique~\cite{PhysRevResearch.6.023050, garttner2017measuring, PhysRevLett.120.040402}. Alternatively, since $\mathcal E_N$ probes a single element of $\hat\varrho_A$ \cite{PhysRevLett.126.210506}, it can be evaluated by quantum-state tomography~\cite{Plodzie2023generation}.}}

{\it Summary and conclusions} ---
\twm{We have investigated a system's dynamics with interactions induces by coupling photons to pairs of atoms. Based on the example of the bosonic Josephson junction immersed into an optical cavity, in which photons are pumped through the cavity mirrors, we have shown how the strength of the atom-atom interaction, resulting from single-photon--single-pair coupling, can be controlled by injecting photons into the cavity.} This allows us to fine-tune the time required for atoms to become entangled without having to rely on the potentially noisy magnetic fields that 
are typically used to control the strength of the atom-atom interaction. \twm{Our analytical results show that the strength of entanglement, as measured by the quantum Fisher information,  depends only on the intensity of the cavity light and not on the precise structure of the photonic state.} The configuration is, therefore, robust to noise from photon losses. 
We analytically predicted the characteristic time-scale for the generation of entanglement, which for realistic parameters can be orders of magnitude shorter than of bare atom-atom collisions.  In turn, the short time scale implies that photon losses during the preparation of the atomic state can be overcome, since strong entanglement can be generated even with large losses. \twm{Additionally, we found the optimal measurement which saturates the quantum Cramer-Rao lower bound. Also, the atoms exhibit strong nonlocal many-body Bell correlations, which demonstrates that the photon-induced interaction can lead to strongly entangled state that can serve for quantum technological purposes as well as for fundamental tests of quantum mechanics~\cite{Schmied_2016,PhysRevLett.126.210506}.}

As a result, many-body entangled states of atoms, with their enormous potential for quantum applications, could be generated very quickly and reliably. In this way, one obstacle to the implementation of quantum solutions can be overcome. 
Our setup can be used in future quantum sensors or other devices where controllable and scalable many-body entanglement is required.

\twm{We have demonstrated that the photon--atom-pair coupling} adds an important ingredient to the techniques of engineering the interactions~\cite{imamoglu2017engineering, davis2020protecting,periwal2021programmable, mivehvar2021cavity}. 
\twm{Due to the vanishing tunneling rate, the dynamics in the protocol do not generate nonclassical light states, see SM~\cite{supp}. 
%% \\ \sks{@Tomasz: Which section of SM you are referring here?}
%% \\ \twc{I have added a comment at the end of section I in SM. Also, we thought of adding a figure with Mandel Q <0 for finite tunneling, in tunneling-dominated regime, right?} \\ 
However, since the photon-assisted atom-atom interaction depends in general on the quantum state of cavity photons, an intriguing direction for future research would be to explore the interplay of atomic and photonic nonclassicality, and how the non-classical nature of photonic states could give rise to novel quantum phases of matter and light~\cite{dmytruk2022controlling}. 
For instance, since the interaction is position-dependent, it may lead to instabilities signalling novel types of self-ordering of atoms driven by correlations~\cite{yamazaki2010submicron, wasak@2013nonlinear, mivehvar2021cavity}, such as supersolidity~\cite{recati2023supersolidity}. Finally, a promising research direction is to explore the potential for generating $n$-body photon-mediated interactions with $n>2$~\cite{luo2024realization}.}

\begin{acknowledgments}
We are grateful to Jean-Philippe Brantut and Helmut Ritsch for their insightful comments on our work.

This research is part of the project No. 2021/43/P/ST2/02911 co-funded by the National Science Centre and the European Union Framework Programme for Research and Innovation Horizon 2020 under the Marie Skłodowska-Curie grant agreement No. 945339.
For the purpose of Open Access, the author has applied a CC-BY public copyright licence to any Author Accepted Manuscript (AAM) version arising from this submission.
J.Ch. acknowledges the support of the National Science Centre, Poland, within the QuantERA II Programme that has received funding from the European Union’s Horizon 2020 research and innovation programme under Grant Agreement No 101017733, Project No. 2021/03/Y/ST2/00195.
\\

%\textit{Data availability:} The data presented in this article is available from~\cite{data}.

\end{acknowledgments}

\newpage

\appendix
\onecolumngrid
\section{Dynamics of the system}

\subsection{Derivation of the effective Hamiltonian}\label{app.ham}

To derive the effective Hamiltonian without the transversal pumping ($\eta_{t} = 0$), we first take the cavity mode function in the form of a standing-wave Gaussian, i.e.,
\begin{equation}\label{5}
   f(\mathbf{r})=  \cos (k y) {e^{-\left(x^2+z^2\right) /(2 \sigma^2)}},
\end{equation}
where $k=\omega_c / c$ is the wave number of the cavity mode, $L$ is the distance between the mirrors, and $\sigma$ is the width of the Gaussian profile in the $x$-$z$ plane.
Furthermore, we switch to a frame rotating with $\omega_{L}$ as follows
\begin{align}
  &\hat a = \Tilde{a}e^{-i\omega_{L}t}\\
  &\psi = \tilde\psi\\
  &\hat\Psi_{\Delta} = \Tilde\psi_{\Delta}e^{-i\omega_{L}t}.
\end{align}
In this scenario, we obtain the set of Heisenberg equations, where the total Hamiltonian is $\hat H = \hat H_A\!+\! \hat H_\Delta\!+\!\hat H_C \!+\! \hat H_{CA} \!+\! \hat H_P$
\begin{subequations}
  \begin{align}\label{7}
    &i\partial_{t}\Tilde{a} = \left[\Tilde{a},\hat{H}-\omega_{L}\Tilde{a}^\dagger\Tilde{a}\right]=
    -\Delta_{c}\Tilde{a} - i\eta +i\Omega_{R}\int d^3rf(r)\left(\Tilde{\Psi}^2_{g}(r)\right)^\dagger\Tilde{\Psi}_{\Delta}(r)\\
    &i\partial_{t}\Tilde{\Psi}_{g}(r) = \left[\Tilde{\Psi}_{g}(r),\hat{H}\right]=\left[\frac{-\nabla^2}{2m}+V(r)\right]\Tilde{\Psi}_{g}(r) + 
    g\Tilde{\Psi}_{g}^\dagger(r)\Tilde{\Psi}_{g}(r)\Tilde{\Psi}_{g}(r)+2i\Omega_{R}f(r)\Tilde{a}^\dagger\Tilde{\Psi}_{g}^\dagger(r)\Tilde{\Psi}_{\Delta}(r)\\
    &i\partial_{t}\Tilde{\Psi}_{\Delta}(r)=\left[\Tilde{\Psi}_{\Delta},\hat{H}-\omega_{L}\int d^3r \Tilde{\Psi}_{\Delta}^\dagger(r)\Tilde{\Psi}_{\Delta}(r)\right]=
    \left[\frac{-\nabla^2}{2M}+V(r)-\Delta_{A}\right]\Tilde{\Psi}_{\Delta}(r) -i\Omega_{R}f(r)\Tilde{a}\Tilde{\Psi}_{g}^2(r)\label{eq.evo.delta}
  \end{align}
\end{subequations}
The above equations of motions can be derived by commuting the corresponding operators with the Hamiltonian $\hat H'$,
where
\begin{equation}
    \hat H' = \hat H - \omega_{L}\Tilde{a}^\dagger\Tilde{a} - \omega_{L}\int d^3r \Tilde{\Psi}_{\Delta}^\dagger(r)\Tilde{\Psi}_{\Delta}(r).
\end{equation}

In the limit, when the atomic detuning $\Delta_A = \omega_L-\omega_B$ is much larger than the other characteristic frequency scales of the system, the population of the molecular state is small and follows the ground state adiabatically. The molecular excited-field operator thus can be eliminated. 
Under these conditions the square bracket in Eq.~\eqref{eq.evo.delta} is dominated by the detuning, and its stationary value can approximate the molecular-state field operator:
\begin{equation}\label{11}
    \Tilde{\Psi}_{\Delta} \simeq -i\frac{\Omega_{R}f(r)}{\Delta_{A}}\Tilde{a}\Tilde{\Psi}_{g}^2(r).
\end{equation}

The resulting dynamics of the photon field and the ground-state atoms are given by
\begin{subequations}
  \begin{align}
    &i\partial_{t}\Tilde{a}=-\Delta_{c}\Tilde{a} - i\eta +\frac{\Omega_{R}^2}{\Delta_{A}}\Tilde{a}\int d^3rf^2(r)\left(\Tilde{\Psi}^2_{g}(r)\right)^\dagger\Tilde{\Psi}_{g}^2(r),\label{eq.a}\\
    &i\partial_{t}\Tilde{\Psi}_{g}(r)=\left[\frac{-\nabla^2}{2m}+V(r)\right]\Tilde{\Psi}_{g}(r) + g\Tilde{\Psi}_{g}^\dagger(r)\Tilde{\Psi}_{g}(r)\Tilde{\Psi}_{g}(r)
    +2\frac{\Omega_{R}^2}{\Delta_{A}}f^2(r)\Tilde{a}^\dagger\Tilde{a}\Tilde{\Psi}_{g}^\dagger(r)\Tilde{\Psi}_{g}^2(r),\label{eq.g}
  \end{align}
\end{subequations}
where we have neglected interactions between molecular state atoms since the population of the molecular state is much smaller than that of the ground state.

The approximate equations of motions given in Eq.~\eqref{eq.a} and \eqref{eq.g} can be derived from the effective Hamiltonian, which reads
\begin{align}
  \hat H_\mathrm{eff}&= -\Delta_{c}\Tilde{a}^\dagger\Tilde{a}-i\eta\left(\Tilde{a}^\dagger-\Tilde{a}\right)+
  \int d^3r \Tilde{\Psi}_{g}^\dagger(r)\left[\frac{-\nabla^2}{2m}+V(r)\right]\Tilde{\Psi}_{g}(r)\nonumber\\
  &+\frac{1}{2}\int d^3r\,g\,\Tilde{\Psi}_{g}^\dagger(r)\Tilde{\Psi}_{g}^\dagger(r)\Tilde{\Psi}_{g}(r)\Tilde{\Psi}_{g}(r)
  + U_{0}\Tilde{a}^\dagger\Tilde{a}\int d^3r f^2(r)\left(\Tilde{\Psi}^2_{g}(r)\right)^\dagger\Tilde{\Psi}_{g}^2,\label{Heff}
\end{align}
where with $U_{0} = \frac{\Omega_{R}^2}{\Delta_{A}}$ we denote the strength of the atom-atom interaction per photon induced by cavity photons.

As a result of the adiabatic elimination of the molecular state, atom-photon interaction stems from the pair-photon scattering by the intermediate process of molecule creation. As a consequence, a new optical potential has appeared with position dependence $f^2(r)$ and an effective amplitude $U_{0}\Tilde{a}^\dagger\Tilde{a}$ which depends on the state of the photon in the cavity and on detuning from molecular state. Since these parameters can be controlled in the experiments, new accessible types of interactions are allowed inside the cavity.

\subsection{Two-mode approximation for bosonic atoms}\label{app.2m}

In order to arrive at a simplified, two-mode description of the system, we assume that the double-well potential from Eq.~(1) defines localized states $w_1(x)$ and $w_2(x)$ centered around the minima of the potential $V_\mathrm{DW}(x)$. The atomic field operator truncated to the lowest lying modes is then approximated as
\begin{equation}\label{psi_2m}
    \Tilde{\Psi}_g(\mathbf{r})  \approx \phi_1(\br) \hat b_1 + \phi_2(\br) \hat b_2,
\end{equation}
where the localized wavefunctions are
\begin{equation}
    \phi_i(\br) = w_i(x) \frac{e^{-\left(y^2+z^2\right) /\left(2 l_H^2\right)}}{\sqrt\pi l_H},
\end{equation}
with $\hat{b}_i$, $i\in \{1,2\}$, the bosonic annihilation operators of the localized modes, and $l_H = \sqrt{{\hbar}/({m\omega_H})}$ is the harmonic oscillator length of the potential in $y$-$z$ plane.

By substituting Eq.~\eqref{psi_2m} into the effective Hamiltonian from Eq.~\eqref{Heff}, and since the double well is symmetric, one arrives at the light-matter Hamiltonian 
\begin{equation}
    \hat{H}_\mathrm{eff} \approx \hat{H}_{L} + \hat{H}_{A} +\hat{H}_{AL},
\end{equation}
where the bare cavity field with pump is described by
\begin{equation}\label{Heff_L}
\hat{H}_{L} = -\Delta_{c}\Tilde{a}^\dagger\Tilde{a}-i\eta\left(\Tilde{a}^\dagger-\Tilde{a}\right)
\end{equation}
The atoms are now described by
\begin{equation}\label{Heff_A}
\hat{H}_A=\epsilon \hat{N}_A-J\left(\hat{b}_1^{\dagger} \hat{b}_2+\hat{b}_2^{\dagger} \hat{b}_1\right)+\frac{U}{2}\left(\hat{b}_1^{\dagger} \hat{b}_1^{\dagger} \hat{b}_1 \hat{b}_1+\hat{b}_2^{\dagger} \hat{b}_2^{\dagger} \hat{b}_2 \hat{b}_2\right)
\end{equation}
where $\hat{N}_{A} = \hat{b}_1^{\dagger}\hat{b}_1 + \hat{b}_2^{\dagger}\hat{b}_2$ is the total number of atoms. The parameter $\epsilon$ is the on-site energy of a single well, $J$ represents the tunneling amplitude, and $U$ is the on-site interaction energy to the bare ground-state atom interactions. 
The dispersive interaction between the atomic pairs and the cavity photons takes the following form:
\begin{equation}\label{Heff_LA}
    \hat{H}_{AL} = W_0 \Tilde{a}^{\dagger}\Tilde{a} \left[\hat{b}_1^{\dagger} \hat{b}_1^{\dagger} \hat{b}_1 \hat{b}_1+\hat{b}_2^{\dagger} \hat{b}_2^{\dagger} \hat{b}_2 \hat{b}_2\right],
\end{equation}
where $W_0$ is the cavity-induced atom-atom coupling strength per photon.
The new parameters are the following:
\begin{subequations}
  \begin{align}
    &\epsilon= \omega_H+\int d x w_j^*(x)\left[-\frac{1}{2 m} \frac{d^2}{d x^2}+V_{\mathrm{DW}}(x)\right] w_j(x)\label{eq.eps}\\
    &J=-\int d x w_1^*(x)\left[-\frac{1}{2 m} \frac{d^2}{d x^2}+V_{\mathrm{DW}}(x)\right] w_2(x),\label{eq.J}\\
    &U=\frac{g}{2 \pi l_H^2} \int d x\left|w_j(x)\right|^4.\label{eq.U}\\
    &W_0 = \frac{ U_{0}\left(1 + e^{-\frac{k^2 l_{H}^2}{2}}\right)}{L \pi^2 \sigma l_{H}^2 \sqrt{2\left(l_H^2 + 2\sigma^2\right)}} \int dx \lvert\omega_{j}(x)\rvert^4 e^\frac{-x^2}{\sigma^2}.\label{eq.W0}
  \end{align}
\end{subequations}

Finally, by means of the Schwinger representation of the angular momentum operators
\begin{align}
  \hat J_x = \frac{b_1^\dagger b_2 + b_2^\dagger b_1}{2},\ \ \hat J_y = \frac{b_1^\dagger b_2 - b_2^\dagger b_1}{2i},\ \ \hat J_z = \frac{b_1^\dagger b_1 - b_2^\dagger b_2}{2},
\end{align}
we obtain the effective BJJ Hamiltonian provided in the main text.
%from Eq.~(5) in the main text.

\subsection{The atomic density operator}\label{app.at}

The first step of the derivation of the atomic density operator is to rotate out the pump term. This is done by means of the displacement operator
\begin{align}
  \hat D(\hat\beta)=e^{\hat\beta(\hat a-\hat a^\dagger)},\ \ \ \hat\beta=\eta\hat\omega^{-1},\ \ \ [\hat\beta,\hat\omega]=0,
\end{align}
where $\hat \omega = -\Delta_c \hat{\mathds1}_A + 2 W_0 \hat J_z^2$ and $\hat{\mathds 1}_A$ is the identity in the atomic subspace.
The action of this operator onto the photon-number operator is
\begin{align}
  \hat D^\dagger(\hat\beta)\hat\omega\hat a^\dagger\hat a\hat D(\hat\beta)=\hat\omega(\hat a^\dagger-\hat\beta)(\hat a-\hat\beta)=\hat\omega\hat a^\dagger\hat a-\hat\omega\hat\beta(\hat a^\dagger+\hat a)+\omega\beta^2
  =\hat\omega\hat a^\dagger\hat a-\eta(\hat a^\dagger+\hat a)+\hat\omega\hat\beta^2.
\end{align}
Hence, by means of this expression, our starting Hamiltonian from Eq.~(5) of the main text (note that we set $J=0$) can be written as
\begin{align}\label{eq.ham.disp}
  \hat H=\hat D^\dagger(\hat\beta)\hat\omega\hat n\hat D(\hat\beta)+U\hat J_z^2-\eta\hat\beta.
\end{align}
Therefore, and since $\hat\beta$ commutes with the remaining operators, we obtain the evolution operator in the form
\begin{align}
  \hat U(t)=\hat D^\dagger(\hat\beta)e^{-i\hat\omega t\hat n}\hat D(\hat\beta)e^{-it(U\hat J_z^2-\eta\hat\beta)}.
\end{align}
We now consider the action of this operator on a generic atom-light input state in the form
\begin{align}
  \hat\varrho(0)=\sum_{nn'=0}^\infty\sum_{mm'=-N/2}^{N/2}\varrho_{nn'}^{mm'}\ketbra{nm}{n'm'},
\end{align}
where $n/n'$ stands for the photon number and $m/m'$ are the eigenvalues of $J_z$ and $\ket{nm}=\ket n_L\otimes\ket m_A$.

The action of the last (atom-only) term of the evolution operator is straightforward, namely
\begin{align}
  e^{-it(U\hat J_z^2-\eta\hat\beta)}\ket{nm}=e^{-it(Um^2-\eta\beta_m)}\ket{nm},\ \ \ \beta_m=\frac\eta{-\Delta_c+2W_0m^2}.
\end{align}
Moreover, also in the first part of the evolution operator $\hat\beta$ can be replaced by $\beta_m$ and $\hat\omega$ by $\omega_m$.
Hence, our goal now is to calculate
\begin{align}
  \hat D^\dagger(\beta_m)e^{-i\omega_m t\hat n}\hat D(\beta_m)\ketbra{n}{n'}\hat D^\dagger(\beta_m)e^{i\omega_m t\hat n}\hat D(\beta_m).
\end{align}
To this end, we notice that 
\begin{align}\label{eq.disp1}
  \ket{\psi_1}\equiv\hat D(\beta_m)\ket{n}=\frac1{\sqrt{n!}}\hat D(\beta_m)(\hat a^\dagger)^n\ket0=\frac1{\sqrt{n!}}\hat D(\beta_m)(\hat a^\dagger)^n\hat D^\dagger(\beta_m)\hat D(\beta_m)\ket0=
  \frac1{\sqrt{n!}}(\hat a^\dagger-\beta_m)^n\ket{\beta_m},
\end{align}
where $\ket{\beta_m}$ stands for a coherent state of light with the amplitude $\beta_m$. Next is the action of the phase-factor that gives
\begin{align}
  \ket{\psi_2}\equiv e^{-i\omega_m t\hat n}\ket{\psi_1}=\frac1{\sqrt{n!}}e^{-i\omega_m t\hat n}(\hat a^\dagger-\beta_m)^ne^{i\omega_m t\hat n}e^{-i\omega_m t\hat n}\ket{\beta_m}=
  \frac1{\sqrt{n!}}(\hat a^\dagger e^{-i\omega_m t}-\beta_m)^n\ket{\beta_me^{-i\omega_m t}}.
\end{align}
Finally, the second displacement operator, in analogy to Eq.~\eqref{eq.disp1}, yields
\begin{align}\label{eq.disp2}
  \ket{\psi_3}\equiv\hat D^\dagger(\beta_m)\ket{\psi_2}&=\frac1{\sqrt{n!}}\hat D^\dagger(\beta_m)(\hat a^\dagger e^{-i\omega_m t}-\beta_m)^n\hat D(\beta_m)\hat D^\dagger(\beta_m)\ket{\beta_me^{-i\omega_m t}}\nonumber\\
  &=\frac1{\sqrt{n!}}((\hat a^\dagger e^{-i\omega_m t}+\beta_m)-\beta_m)^n\hat D(-\beta_m)\hat D(\beta_me^{-i\omega_m t})\ket0.
\end{align}
Using the property $\hat D(x)\hat D(y)=\hat D(x+y)e^{\frac12(xy^*-x^*y)}$, we obtain the final expression
\begin{align}\label{eq.psi3}
  \ket{\psi_3}=\frac1{\sqrt{n!}}e^{-i\beta^2_m\sin(\omega_mt)}((\hat a^\dagger e^{-i\omega_m t}+\beta_m)-\beta_m)^n\ket{\beta_me^{-i\omega_m t}-\beta_m}.
\end{align}
In order to simplify the above expression, we introduce 
\begin{equation}\label{gamma-m}
    \gamma_m(t) \equiv \beta_m \big(e^{-i\omega_m t}-1\big),
\end{equation} 
which then results in
\begin{align}
  \ket{\psi_3}=\frac1{\sqrt{n!}}e^{-i\beta^2_m\sin(\omega_mt)}(\hat a^\dagger e^{-i\omega_m t}+\gamma_m)^n\ket{\gamma_m}.
\end{align}
In the next step, we observe that
\begin{align}
  (\hat a^\dagger e^{-i\omega_m t}+\gamma_m)^n\ket{\gamma_m}=\hat D(\gamma_m)\hat D^\dagger(\gamma_m)(\hat a^\dagger e^{-i\omega_m t}+\gamma_m)^n\hat D(\gamma_m)\ket0=
  \hat D(\gamma_m)((\hat a^\dagger+\gamma^*) e^{-i\omega_m t}+\gamma_m)^n\ket0.
\end{align}
Using the identity
\begin{align}
  \gamma^*e^{-i\omega_m t}+\gamma_m=\beta_m-\beta_me^{-i\omega_mt}-\beta_m+\beta_me^{-i\omega_mt}=0
\end{align}
we have
\begin{align}
  (\hat a^\dagger e^{-i\omega_m t}+\gamma_m)^n\ket{\gamma_m}=\hat D(\gamma_m)\hat D^\dagger(\gamma_m)(\hat a^\dagger e^{-i\omega_m t}+\gamma_m)^n\hat D(\gamma_m)\ket0=
  \hat D(\gamma_m)(\hat a^\dagger)^n e^{-i\omega_m nt}\ket0.
\end{align}
Hence
\begin{align}
  \ket{\psi_3}=e^{-i\beta^2_m\sin(\omega_mt)}e^{-i\omega_m nt}\hat D(\gamma_m)\ket n.
\end{align}
This way, after all these manipulations, we arrive at the complete expression for the atom-photon density matrix
\begin{align}\label{eq.final}
  \hat\varrho(t)=\hat U(t)\hat\varrho(0)\hat U^{\dagger}(t)&=
  \sum_{nn'=0}^\infty\sum_{mm'=-N/2}^{N/2}\varrho_{nn'}^{mm'}e^{-i(\beta^2_m\sin(\omega_mt)-\beta^2_{m'}\sin(\omega_{m'}t))}e^{-i(\omega_m n-\omega_{m'}m')t}e^{-iU(m^2-m'^2)t}\times\nonumber\\
  &\times\,e^{-i\eta t(\beta_m-\beta_{m'})}\hat D(\gamma_m)\ketbra{nm}{n'm'}\hat D^\dagger(\gamma_m).
\end{align}
The results of the main text uses this atomic density operator, which is obtained by assuming that initially atoms and photons form a separable state pure state, i.e., 
\begin{align}
    \varrho_{nn'}^{mm'}=C_mC^*_{m'}d_nd^*_{n'}
\end{align}
and tracing-out the photonic degrees of freedom, which gives
\begin{align}
  \hat\varrho_A(t)=\sum_{n=0}^\infty\braket n{\psi(t)}\braket{\psi(t)}n=\sum_{mm'}C_m(t)C^*_{m'}(t)\ketbra m{m'}e^{-\frac12(\modsq{\gamma_m}+\modsq{\gamma_{m'}})}e^{\gamma_m\gamma^*_{m'}},
\end{align}
with
\begin{align}
    C_m(t)=C_me^{-i\beta^2_m\sin(\omega_mt)}e^{-iUm^2t}e^{-i\eta t\beta_m},
\end{align}
and $\gamma_m(t)$ is given in Eq.~\eqref{gamma-m}.

\section{Derivation of the QCRB}\label{app.qcrb}

\subsection{Verifying the $\tr{(\hat\varrho_A(t)\hat J_y)^2}=0$ condition}\label{app.ver}

For the generator $\hat J_y$ of transformation, the QFI is given by
\begin{equation}
    F_Q = 2 \sum_{i,j} \frac{(\lambda_i - \lambda_j)^2}{\lambda_i + \lambda_j} |\langle \Psi_i|\hat J_y | \Psi_j\rangle|^2,
\end{equation}
where the eigen-decomposition of the atomic state is $\hat \varrho_A = \sum_i \lambda_i |\Psi_i\rangle \langle \Psi_i|$.
In general, the following inequality holds:
\begin{equation}\label{qfi_variance_bound}
    F_Q \leqslant 4 \big( \mathrm{Tr}[ \hat\varrho_A \hat J_y^2 ] - \mathrm{Tr}[ \hat\varrho_A \hat J_y ]^2\big).
\end{equation}
According to Ref.~\cite{braunstein1994statistical}, the sufficient and necessary condition for the equality is 
\begin{equation} \label{condition_ij}
    \lambda_i \lambda_j |\langle{\Psi_i}| \hat J_y - \av{\hat J_y} | \Psi_j\rangle |^2 = 0
\end{equation}
for all $i$ and $j$; here $\av{\hat J_y} = \tr{ \varrho_A \hat J_y}$. 
Now, we observe that in Eq.~\eqref{condition_ij}, for all $i$ and $j$ the terms are non-negative. Therefore, after summation we obtain an equvalent statement, that Eq.~\eqref{condition_ij} is satisfied if and only if
\begin{equation}
    \sum_{i,j} \lambda_i \lambda_j |\langle{\Psi_i}| \hat J_y - \av{\hat J_y} | \Psi_j\rangle |^2 = 0
\end{equation}
is satisfied. Using the hermiticity of the operator $\hat J_y - \av{\hat J_y}$ and the decomposition of the state $\hat\varrho_A$, the sum can be rewritten as
\begin{align} \label{qfi_condition}
  \tr{(\hat\varrho_A(t)\hat J_y)^2}=0.
\end{align}
This is the condition for the equality in Eq.~\eqref{qfi_variance_bound}.

Below, we demonstrate that the condition from Eq.~\eqref{qfi_condition} is satisfied. With the symmetry of the initial state, we obtain $\av{\hat J_y(t)} = 0$. These two observations imply that that the QCRB from Eq.~(5) simplifies to the formula
\begin{align}
  F_Q=4\tr{\hat\varrho_A(t)\hat J_y^2}.
\end{align}

To proceed with showing Eq.~\eqref{qfi_condition}, we use the expression for the atomic density matrix
\begin{align}\label{eq.corrected}
  \hat\varrho_A(t)=\sum_{mm'=-N/2}^{N/2}C_mC_{m'}^*e^{-i(\beta^2_m\sin(\omega_mt)-\beta^2_{m'}\sin(\omega_{m'}t))}e^{-iU(m^2-m'^2)t}e^{+i\eta t(\beta_m-\beta_{m'})}e^{-\frac12(\modsq{\gamma_m}+\modsq{\gamma_{m'}})}
  e^{\gamma_m\gamma_{m'}^*}\ketbra{m}{m'},
\end{align}
and write it in a compact form
\begin{align}
  &\hat\varrho_A(t)=\sum_{mm'=-N/2}^{N/2}C_mC_{m'}^*e^{-i\varphi_{mm'}}e^{\Gamma_{mm'}}\ketbra{m}{m'},\ \ \ \mathrm{with}\ \ \ \\
  &\varphi_{mm'}=\beta^2_m\sin(\omega_mt)-\beta^2_{m'}\sin(\omega_{m'}t)+U(m^2-m'^2)t-\eta t(\beta_m-\beta_{m'}),\ \ \ \\
  &\Gamma_{mm'}=-\frac12(\modsq{\gamma_m}+\modsq{\gamma_{m'}})+\gamma_m\gamma_{m'}^*.
\end{align}
The action of the $\hat J_y$ operator on the density matrix results in
\begin{align}
  \hat J_y\hat\varrho_A(t)=\frac1{2i}\sum_{mm'=-N/2}^{N/2}C_mC_{m'}^*e^{-i\varphi_{mm'}}e^{\Gamma_{mm'}}\left(\sqrt{\frac N2+m+1}\sqrt{\frac N2-m}\ket{m+1}
  -\sqrt{\frac N2+m}\sqrt{\frac N2-m+1}\ket{m-1}\right)\bra{m'}
\end{align}
Hence, the square of this operator yields
\begin{align}
  (\hat J_y\hat\varrho_A(t))^2&=-\frac14\sum_{mm';nn'=-N/2}^{N/2}C_mC_{m'}^*C_nC_{n'}^*e^{-i(\varphi_{mm'}+\varphi_{nn'})}e^{\Gamma_{mm'}+\Gamma_{nn'}}\times\nonumber\\
  &\left(\sqrt{\frac N2+m+1}\sqrt{\frac N2-m}\ket{m+1}-\sqrt{\frac N2+m}\sqrt{\frac N2-m+1}\ket{m-1}\right)\bra{m'}\times \nonumber\\
  &\left(\sqrt{\frac N2+n+1}\sqrt{\frac N2-n}\ket{n+1}-\sqrt{\frac N2+n}\sqrt{\frac N2-n+1}\ket{n-1}\right)\bra{n'}.
\end{align}
We now perform the inner multiplication and get rid of the sum over $m'$, to obtain
\begin{align}
  (\hat J_y\hat\varrho_A(t))^2&=-\frac14\sum_{m;nn'=-N/2}^{N/2}\\
  &\Bigg[C_mC_{n+1}^*C_nC_{n'}^*e^{-i(\varphi_{m(n+1)}+\varphi_{nn'})}e^{\Gamma_{m(n+1)}+\Gamma_{nn'}}\times\nonumber\\
    &\Bigg(\sqrt{\frac N2+m+1}\sqrt{\frac N2-m}\sqrt{\frac N2+n+1}\sqrt{\frac N2-n}\ketbra{m+1}{n'}+\\
    &-\sqrt{\frac N2+m}\sqrt{\frac N2-m+1}\sqrt{\frac N2+n+1}\sqrt{\frac N2-n}\ketbra{m-1}{n'}\Bigg)+\\
    &-C_mC_{n-1}^*C_nC_{n'}^*e^{-i(\varphi_{m(n-1)}+\varphi_{nn'})}e^{\Gamma_{m(n-1)}+\Gamma_{nn'}}\times\\
    &\Bigg(\sqrt{\frac N2+m+1}\sqrt{\frac N2-m}\sqrt{\frac N2+n}\sqrt{\frac N2-n+1}\ketbra{m+1}{n'}+\\
    &-\sqrt{\frac N2+m}\sqrt{\frac N2-m+1}\sqrt{\frac N2+n}\sqrt{\frac N2-n+1}\ketbra{m-1}{n'}\Bigg)\Bigg].
\end{align}
In the next step, we evaluate the trace of this matrix
\begin{subequations}
  \begin{align}
    &\tr{(\hat J_y\hat\varrho_A(t))^2}=-\frac14\sum_{mn=-N/2}^{N/2}\nonumber\\
    &\Bigg[C_mC_{n+1}^*C_nC_{m+1}^*e^{-i(\varphi_{m(n+1)}+\varphi_{n(m+1)})}e^{\Gamma_{m(n+1)}+\Gamma_{n(m+1)}}\sqrt{\frac N2+m+1}\sqrt{\frac N2-m}\sqrt{\frac N2+n+1}\sqrt{\frac N2-n}\label{eq.l1}\\
      &-C_mC_{n+1}^*C_nC_{m-1}^*e^{-i(\varphi_{m(n+1)}+\varphi_{n(m-1)})}e^{\Gamma_{m(n+1)}+\Gamma_{n(m-1)}}\sqrt{\frac N2+m}\sqrt{\frac N2-m+1}\sqrt{\frac N2+n+1}\sqrt{\frac N2-n}\label{eq.l2}\\
      &-C_mC_{n-1}^*C_nC_{m+1}^*e^{-i(\varphi_{m(n-1)}+\varphi_{n(m+1)})}e^{\Gamma_{m(n-1)}+\Gamma_{n(m+1)}}\sqrt{\frac N2+m+1}\sqrt{\frac N2-m}\sqrt{\frac N2+n}\sqrt{\frac N2-n+1}\label{eq.l3}\\
      &+C_mC_{n-1}^*C_nC_{m-1}^*e^{-i(\varphi_{m(n-1)}+\varphi_{n(m-1)})}e^{\Gamma_{m(n-1)}+\Gamma_{n(m-1)}}\sqrt{\frac N2+m}\sqrt{\frac N2-m+1}\sqrt{\frac N2+n}\sqrt{\frac N2-n+1}\Bigg]\label{eq.l4}.
  \end{align}
\end{subequations}
We now notice that if we shift the index $m$ in line~\eqref{eq.l2}, we obtain~\eqref{eq.l1} (with a reversed sign), for as long as $C_m=C^*_m$. The same argument applies to the pair of
lines~\eqref{eq.l3} and~\eqref{eq.l4}. Hence the whole expression vanishes when the initial amplitudes $C_m\in\mathds R\ \forall\ m\in\{-N/2\ldots N/2\}$, which is the case of the coefficients from
the main text. 

A similar argument shows that this condition also holds for $\eta=0$ and a coherent initial state of photons.

\subsection{The QCRB for the $\eta=0$ case}

When Eq.~\eqref{qfi_condition} is satisfied and $\av{\hat J_y(t)}=0$, which is our case, we can use the
following formula for the QFI
\begin{align}
  F_Q=4\Delta^2\hat J_y.
\end{align}
Using the photonic coherent state and the spin-coherent atomic state
\begin{align}\label{eq.fq.nopump}
  F_q=&4\av{\hat J_y^2}=\frac12N(N+1)-2\re{I},
\end{align}
where
\begin{align}
  I=\sum_mC_mC_{m-2}^*e^{-iU(m^2-(m-2)^{2})t}e^{\modsq\alpha{e^{-i(\omega_m-\omega_{m-2})t}-1}}\sqrt{\left(\frac N2+m+1\right)\left(\frac N2+m+2\right)\left(\frac N2-m\right)\left(\frac N2-m-1\right)}.
\end{align}
We now {\bf assume that $N$ is sufficiently large} to replace the summation over $m$ by an integral. 
For as long as the only quickly changing terms are the phase-factors, the
term under the square-root can be assumed to be slowly varying, hence,
\begin{align}
  I\simeq\sqrt{\frac2{\pi N}}\int_{-\infty}^\infty\!\! dm\,e^{-\frac2Nm^2}e^{4iUt(m+1)}e^{|\alpha|^2\left[e^{i8W_0t(m+1)}-1\right]}\left(\frac{N^2}4-m^2\right).
\end{align}
In the next step, we make an additional assumption, that the times at which this integral is considered are sufficiently short so that $e^{4iW_0t(m+1)}-1\simeq4iW_0t(m+1)$. 
To this end, we notice that the Gaussian function yields the maximal range of $m$'s to be of the order of $\sqrt N$, hence ``short times'' implies that {\bf the condition $4W_0t\sqrt N\ll1$}
must be fulfilled. In this case, we obtain a simple Gaussian integral
\begin{align}
  I=\sqrt{\frac2{\pi N}}\int_{-\infty}^\infty\!\! dm\,e^{-m^2\left(\frac2N+32\bar n(W_0t)^2\right)}e^{4iUt(m+1)}e^{\bar n4iW_0t(m+1)-32\bar n(W_0t)^2(2m+1)}\left(\frac{N^2}4-m^2\right),
\end{align}
where  $\bar n=|\alpha|^2$. This can be written as
\begin{align}
  I=\sqrt{\frac2{\pi N}}e^{i\alpha}\int_{-\infty}^\infty\!\! dm\,e^{-m^2\sigma}e^{i\beta m}\left(\frac{N^2}4-m^2\right),
\end{align}
where
\begin{align}
  \sigma=\frac2N+32\bar n(W_0t)^2,\ \ \ \alpha=a+ib,\ \ \ \beta=-a-2ib,\ \ \ a=-8W_0t\bar n-4Ut,\ \ \ b=32\bar n(W_0t)^2.
\end{align}
The outcome of this Gaussian integral is
\begin{align}
  I=\sqrt{\frac{2}{N\sigma}}e^{-\frac{\beta^2}4\sigma}e^{i\alpha}\left(\frac{\beta^2}{4\sigma^2}+\frac{N^2}4-\frac1{2\sigma}\right).
\end{align}
Hence the QFI from Eq.~\eqref{eq.fq.nopump} is approximated by
\begin{align}\label{eq.int.app}
  F_Q\simeq\frac12N(N+1)-2\sqrt{\frac2{\sigma N}}e^{-\frac1{4\sigma}(a^2-4b^2)}e^{-b}\left[\cos\left(a-\frac{ab}\sigma\right)\left(\frac{N^2}4-\frac1{2\sigma}+\frac{a^2-4b^2}{4\sigma^2}\right)-
    \frac{ab}\sigma\sin\left(a-\frac{ab}\sigma\right)\right].
\end{align}
It is now our goal to identify the dominant terms. 
In order to observe the positive impact of the cavity photons, {\bf it must hold that $W_0\bar n\gg U$}, so $a\simeq-4W_0t\bar n$. 
Moreover {\bf when the mean number of photons is large, $\bar n\gg1$}, then $b\simeq a/{\bar n}$ and can be neglected as compared to $a$.
In such case, the QFI can be approximated as follows
\begin{align}
  F_Q\simeq\frac12N(N+1)-2\sqrt{\frac2{\sigma N}}e^{-\frac1{4\sigma}a^2}e^{-b}\left[\cos\left(a-\frac{ab}\sigma\right)\left(\frac{N^2}4-\frac1{2\sigma}+\frac{a^2}{4\sigma^2}\right)-
    \frac{ab}\sigma\sin\left(a-\frac{ab}\sigma\right)\right].
\end{align}
We now focus on the exponents. We have
\begin{align}
  \frac1{4\sigma}a^2=\frac{16(W_0\bar nt)^2}{\frac2N+32\bar n(W_0t)^2}=\frac{8(W_0\bar nt)^2N}{1+16\frac{(W_0\bar nt)^2N}{\bar n}}.
\end{align}
The characteristic time-scale of the numerator is $\tau_c=\frac{1}{2W_0\bar n\sqrt N}$ and is $\bar n$ shorter than of the denominator, hence, at the times when the exponent is non-negligible, the denominator can be taken as constant, giving
\begin{align}
  \frac1{4\sigma}a^2\simeq2\frac{t^2}{\tau_c^2}.
\end{align}
At these times $b\ll1$, and, hence, another step of the approximation is 
\begin{align}
  F_Q\simeq\frac12N(N+1)-2\sqrt{\frac2{\sigma N}}e^{-\frac1{4\sigma}a^2}\left[\cos\left(a-\frac{ab}\sigma\right)\left(\frac{N^2}4-\frac1{2\sigma}+\frac{a^2}{4\sigma^2}\right)-
    \frac{ab}\sigma\sin\left(a-\frac{ab}\sigma\right)\right].
\end{align}
Also, when $t\simeq\tau_c$, it holds that $a\ll1$ and $\sigma\simeq 2/N$, hence, we obtain the final expression
\begin{align}
  F_Q\simeq\frac12N(N+1)-e^{-2\frac{t^2}{\tau_c^2}}\left(\frac{N^2}2-\frac N2+4\frac{t^2}{\tau_c^2}\frac N2\right).
\end{align}
This last term in the parenthesis can be safely neglected, as for short times it is small, while when $\tau\sim\tau_c$, the Gaussian nullifies the whole term inside the parenthesis.
Therefore, under the above assumptions, the QFI is
\begin{align}
  F_Q\simeq\frac12N\left(1+f(t)\right)+\frac12N^2\left(1-f(t)\right),\ \ \ f(t)=e^{-2\frac{t^2}{\tau_c^2}},\ \ \ \tau_c=\frac1{2\sqrt{N}W_0\bar n}.
\end{align}

\subsection{The QCRB for the $\eta\neq0$ case}
\label{app-coh-qfi}

Employing the same approach as above but with photons initially in a vacuum state we obtain again
\begin{align}
  F_q=&4\av{\hat J_y^2}=\frac12N(N+1)-2\re{I},
\end{align}
with the term $I$ given by
\begin{align}
  I&=\sum_mC_mC_{m+2}e^{-iU(m^2-(m+2)^{2})t}e^{-i(\beta^2_m\sin(\omega_mt)-\beta^2_{m+2}\sin(\omega_{m+2}t))}\\
  &\times e^{i\eta t(\beta_m-\beta_{m+2})}e^{-\frac12(\modsq{\gamma_m}+\modsq{\gamma_{m+2}})}e^{\gamma_m\gamma^*_{m+2}}
  \sqrt{\bigg(\frac N2+m+1\bigg)\bigg(\frac N2+m+2\bigg)\bigg(\frac N2-m\bigg)\bigg(\frac N2-m-1\bigg)}\Big].\nonumber
\end{align}
The {\bf condition for short times requires  now that $\Delta_ct\ll1$ is satisfied}.  
In this case we can expand the trigonometric functions and to the leading orded we obtain:
\begin{subequations}
  \begin{align}
    &e^{-\frac12(\modsq{\gamma_m}+\modsq{\gamma_{m+2}})}e^{\gamma_m\gamma^*_{m+2}}\simeq e^{i\eta^2W_0t^34(m+1)},\\
    &e^{i\eta t(\beta_m-\beta_{m+2})}=e^{-i\frac13\eta^2W_0t^34(m+1)}.
  \end{align}
\end{subequations}
The approximate expression for the integral is then
\begin{align}
  I=\sqrt{\frac2{\pi N}}\int_{-\infty}^\infty\!\! dm\,e^{-\frac2Nm^2}e^{4iUt(m+1)}e^{i\frac83\eta^2W_0(m+1)t^3}\left(\frac{N^2}4-m^2\right).
\end{align}
This, in turn, yields the QFI equal to
\begin{align}
  F_Q\simeq\frac12N(N+1)-e^{-\frac{\chi(t)^2N}8}\left(\frac{N^2}2(1+\frac{\chi(t)^2}4)-\frac N2\right)\cos(\chi(t)),
\end{align}
where $\chi=4(Ut+\frac23\eta^2W_0t^3)$. Note that the Gaussian drops to zero at times much faster then the time-variance of other terms, because it has the $N$ coefficient in the exponent.
Hence, we arrive at the final expression for the QFI
\begin{align}
  F_Q\simeq\frac12N\left(1+f(t)\right)+\frac12N^2\left(1-f(t)\right),\ \ \ f(t)=e^{-2N(Ut+\frac23\eta^2W_0t^3)^2}.
\end{align}
When {\bf the mean number of photons $\bar n(t)=\eta^2t^2$ is sufficiently large,  i.e., $W_0 \bar n(t) \gg U$}, we recover the formula from the main text.

\section{Parameters}

Our effective Hamiltonian for the photon-pair coupling is:
\begin{equation}
  \hat H_{CA} = -i \Omega_R \int d^3r f(\br) \hat\Psi_\Delta^\dag(\br) \hat\Psi_g^2(\br) \hat a + \mathrm{H.c.},
\end{equation}
where $\Omega_R$ is the strength of the light-matter coupling and $f(\br)$ is the cavity photon mode function, $\hat\Psi_\Delta$ ($\hat\Psi_g$) is the molecule (ground-state atom) annihilation field operator, and $\hat a$ is the annihilation operator of the single-mode cavity photon.

In Ref.~\cite{konishi2021universal}, the microscopic Hamiltonian that is used for the description of the single-photon--single-pair interaction has the following structure:
\begin{equation}
    \hat H_\mathrm{pair} = -i \frac{\Omega_0}{2}  \int d^3R \int d^3r f(\mathbf{R}) \phi_m(\br)
    \hat\Psi_\Delta^\dag(\mathbf{R}) 
    \hat\Psi_g\bigg(\mathbf{R} - \frac{\br}{2}\bigg)\hat\Psi_g\bigg(\mathbf{R} + \frac{\br}{2}\bigg) \hat a + \mathrm{H.c.},
\end{equation}
where $\phi_m(\mathbf{r})$ is a wavefunction that describes the relative motion of atoms in the target molecule state. Here, $\Omega_0$ is the single-photon--single-pair Rabi coupling.
Assuming that the dependence on $\br$ of the field operators changes weakly on the characteristic scales of the problem, we approximate:
\begin{equation}
    \hat H_\mathrm{pair} \approx -i \frac{\Omega_0}{2}\xi  \int d^3R f(\mathbf{R}) 
    \Psi_\Delta^\dag(\mathbf{R}) 
    \Psi_g(\mathbf{R})\Psi_g(\mathbf{R}) a + \mathrm{H.c.},
\end{equation}
where
\begin{equation}
    \xi =  \int d^3r  \phi_m(\br),
\end{equation}
which matches Eq.~(4) with
\begin{equation}
    \Omega_R = \frac{\Omega_0 \xi}{2}.
\end{equation}
We note that the units of $\Omega_R$ are $[\mathrm{frequency}][\mathrm{length}]^{3/2}$ and $\Omega_0 \sim [\mathrm{frequency}]$ with $\hbar=1$.
Therefore, we find that, after adiabatic elimination of the molecular state in the low saturation regime, i.e., when the number of molecules is vanishingly small, the atomic interaction Hamiltonian takes the form:
\begin{equation}\label{ham_gu}
    \hat H_\mathrm{int} = \int d^3r \bigg(\frac{g}{2} + U_0 a^\dag a f^2(\br)\bigg) \Psi_g^\dag(\br) \Psi_g^\dag(\br) \Psi_g(\br)\Psi_g(\br), 
\end{equation}
where the cavity-assisted atom-atom interaction strength per photon is
\begin{equation}
    U_0 = \frac{\Omega_R^2}{\Delta_A},
\end{equation}
where $\Delta_A = \omega_L - \omega_B$ is the detuning of the pump laser with respect to the molecular state.

Critical, therefore, is the ratio of $U_0$ and $g$. In order to estimate its value, we make a simple assumption about the molecular wavefunction, i.e., $\phi_m(\br) \propto \exp(- |\br|^2 / R_0^2)/|\br|$, where $R_0$ is the characteristic length scale of the molecular state.
In such a case, we obtain $\xi = 2 \sqrt{2 \pi} R_0^{3/2}$.

Finally, for an order-of magnitude estimation, we take $\Omega_0 \sim 2\pi \times 750$ kHz, which is an order of magnitude reported in Ref.~\cite{konishi2021universal}, detuning $\Delta_a \sim 2\pi \times 2$ MHz, atomic mass $m \sim 87$ u, where  u is the atomic mass unit, the scattering length $a \sim 100 a_0$, with $a_0$ the Bohr radius. We are assuming that the $\Delta_a$ is much larger than the decay rate of the molecular state. For the scale $R_0 \sim R_C$, where the Condon point of the molecular state $R_C\sim 250 a_0$, as reported in Ref.~\cite{konishi2021universal}. For these parameters, we obtain $U_0/g \sim 0.5$. The cavity-assisted atom-atom interaction strength per cavity photon, $U_0$ is, thus, of the order of the bare coupling $g$.

For the double-well geometry,  $\hat\Psi_g(\br) = \phi_1(\br) \hat b_1 + \phi_2(\br) \hat b_2 $, where $\hat b_i$, with $i=1,2$, are the modes that are centred in the tight traps. The couplings that enter into the two-mode Hamiltonian are:
\begin{equation}
    U = g \int d^3 r |\phi(\br)|^4
\end{equation}
for the bare atom-atom interaction, and
\begin{equation}
    W_0 = U_0 \int d^3 r [f(\br)]^2 |\phi(\br)|^4
\end{equation}
for the cavity-induced atom-atom interaction strength. We see that the ratio
\begin{equation} \label{ratio}
    \frac{W_0}{U}  = \frac{U_0}{g} \times \frac{\int d^3 r [f(\br)]^2 |\phi(\br)|^4}{\int d^3 r |\phi(\br)|^4}
\end{equation}
is of the order $U_0/g$ times a factor that takes into account the mode function of the cavity.
Assuming that the tight atomic traps are centered in the region $f\sim1$, we obtain $W_0/U \sim U_0 /g \sim 1$. For lighter atoms (like ${}^7$Li) we may have $W_0/g \sim 0.1 \ll 1$ if we set the scattering length $a \sim 25 a_0$. We note that the value $U_0/g$ strongly depends on $R_C$, and for small $R_C$ one may have $U_0/g \ll 1$. In such a case, more photons or stronger pumps are needed in order to overcome the bare atom-atom interaction strength in Eq.~\eqref{ham_gu}.

\section{Dynamics with coherent state of cavity mode}

In this section, we focus on the scenario where the cavity contains a coherent state of photons with amplitude $\alpha$ and no external pump is applied ( $\eta = 0$ ). This case is analogous to the optical Feshbach resonance in free space with a coherent beam~\cite{chin2010feshbach}, which can lead to dynamical instabilities in Bose-Einstein condensates~\cite{wasak@2013nonlinear}.

Initially, the input atomic and photonic states are separable. The evolved atomic state at time $t$, after tracing out the photonic degrees of freedom, is given by
\begin{align}\label{eq:rho_a}
  \hat{\varrho}_A(t) = \sum_{m,m'} C_m C_{m'} e^{-iU(m^2 - m'^2)t} u_{mm'}(t) \ketbra{m}{m'},
\end{align}
where
\begin{align}\label{eq:u_a}
  u_{mm'}(t) = \exp\left\{ |\alpha|^2 \left[ e^{-i(\omega_m - \omega_{m'}) t} - 1 \right] \right\}
\end{align}
with $\omega_m=-\Delta_c+2W_0m^2$, $\beta_m=\eta/\omega_m$ and $\gamma_m=\beta_m(e^{-i\omega_m t}-1)$. We note that $u_{mm'}(0) = 1$.
Under similar assumptions as that of the pumped case, we can derive an approximate analytical expression for the QFI of the atomic subsystem, see Sec.~\ref{app-coh-qfi}:
\begin{align}\label{eq:QFI_approx}
  F_Q(t) \simeq \frac{1}{2} N \left[ 1 + f(t) \right] + \frac{1}{2} N^2 \left[ 1 - f(t) \right],
\end{align}
where we have the dynamical crossover function given by:
\begin{align}\label{eq:f_a}
  f(t) = \exp\left[ -2 \left( \frac{t}{\tau_{\mathbf{C}}} \right)^2 \right], \quad \tau_{\mathbf{C}} = \frac{1}{2\sqrt{N} W_0 |\alpha|^2}.
\end{align}

\begin{figure}[t!]
  \centering
        \includegraphics[width=1.0\linewidth]{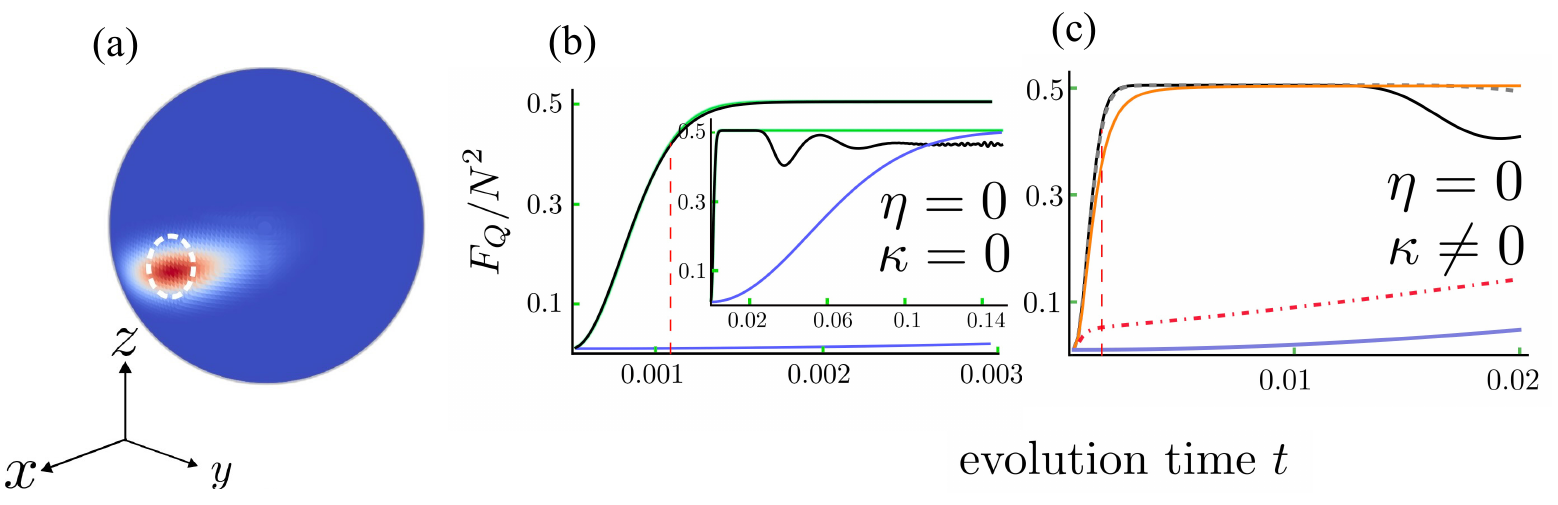}
        \caption{
        {\bf (a)}: 
        Husimi Q-function without cavity light-matter coupling  after time $t = 1/(4\sqrt{N}) U^{-1}$ at which the spin-squeezing  is already visible in the regular OAT scheme. The dashed white line corresponds to the  initial state distribution.
        {\bf (b)}:
        The dynamics of the QFI for lossless ($\kappa=0$) and {\bf (c)}: The dynamics  for lossy ($\kappa\neq0$) cases ; the time is in units of $U^{-1}$. 
        The input coherent state contains $|\alpha|^2=40$ photons. The solid black lines show the QFI for $\kappa=0$  using the atomic density matrix from Eq.~\eqref{eq:u_a} with $u_{mm'}(t)$.
        The solid blue lines show the QFI for the OAT method (no photons).
        The solid green lines are the approximations from Eq.~\eqref{eq:QFI_approx}.
        The effect of losses on the QFI  is for: $\kappa \tau_{\mathbf{C}} = 0.5 \times \left( 0.1 \, \text{(dashed grey)}, 1 \, \text{(orange)}, 10 \, \text{(dot-dashed red)} \right)$
        The vertical red dashed lines denote: $\tau_{\mathbf{C}}=0.0025$.
        The system parameters are: $U=W_0=\Delta_c=1$, $N=100$ atoms.}
    \label{fig.coherent_total}
\end{figure}

For the coherent state of photons the timescale $\tau_{\mathbf{C}} \propto N^{-1/2}$ is similar to standard OAT scheme.
It is important to note that, without the cavity light-matter coupling, we obtain spin-squeezed states, as evidenced in Fig \ref{fig.coherent_total}$(\bf{a})$. However, when the cavity is included, i.e., with light-matter coupling, the system transitions into a non-Gaussian regime, as discussed in the main text of our manuscript for the pumped case. We have taken into account three orders of magnitude of looses $\kappa \tau_{\mathbf{C}} = 0.5 \times (0.1,1,10)$
%\twc{Factor of 2 not taken into account?}.

As that of the pumped case our results in Figs. \ref{fig.coherent_total}$(\bf{b})$ and $(\bf{c})$ indicate that as long as  $\kappa^{-1} \gtrsim \tau_{\mathbf{C}}$, the QFI remains largely unaffected by photon losses, demonstrating the robustness of the entanglement generation against decoherence in the photonic subspace. Even for significant losses ($\kappa^{-1} \sim \tau_{\mathbf{C}}$), the atomic entanglement builds up efficiently, albeit on a modified timescale.

%\twc{I think we can remove this part below as it repeats what is in the main text.}

%\tws{These findings highlight that the QFI is determined primarily by the average number of photons rather than photonic coherences. Consequently, the cavity-assisted One-Axis Twisting procedure generates strong many-body entanglement that is resilient to photon losses. This robustness is crucial for experimental implementations where photon losses are inevitable.}

%\tws{In summary, we have shown that the light-mediated coupling of atoms to a coherent photon field can drastically accelerate the generation of strong entanglement compared to the standard OAT scheme. The collective coupling enhances the effective interaction strength, leading to a rapid build-up of Quantum Fisher Information towards the Heisenberg limit. Importantly, this acceleration occurs even in the presence of photon losses, making the scheme highly advantageous for practical applications in quantum metrology.}

\section{Entanglement After Pump}

In this Section, we analyze the dynamics of the QFI $F_Q$ after the pump is switched off at finite time $t$. We carry out numerical simulations in the regime where the cavity loss rate $\kappa$ is on the order of the timescale $\tau_Q$, i.e., $\kappa \tau_Q \approx 1$.

\begin{figure}[t!]
     \centering
     \includegraphics[width=0.49\linewidth]{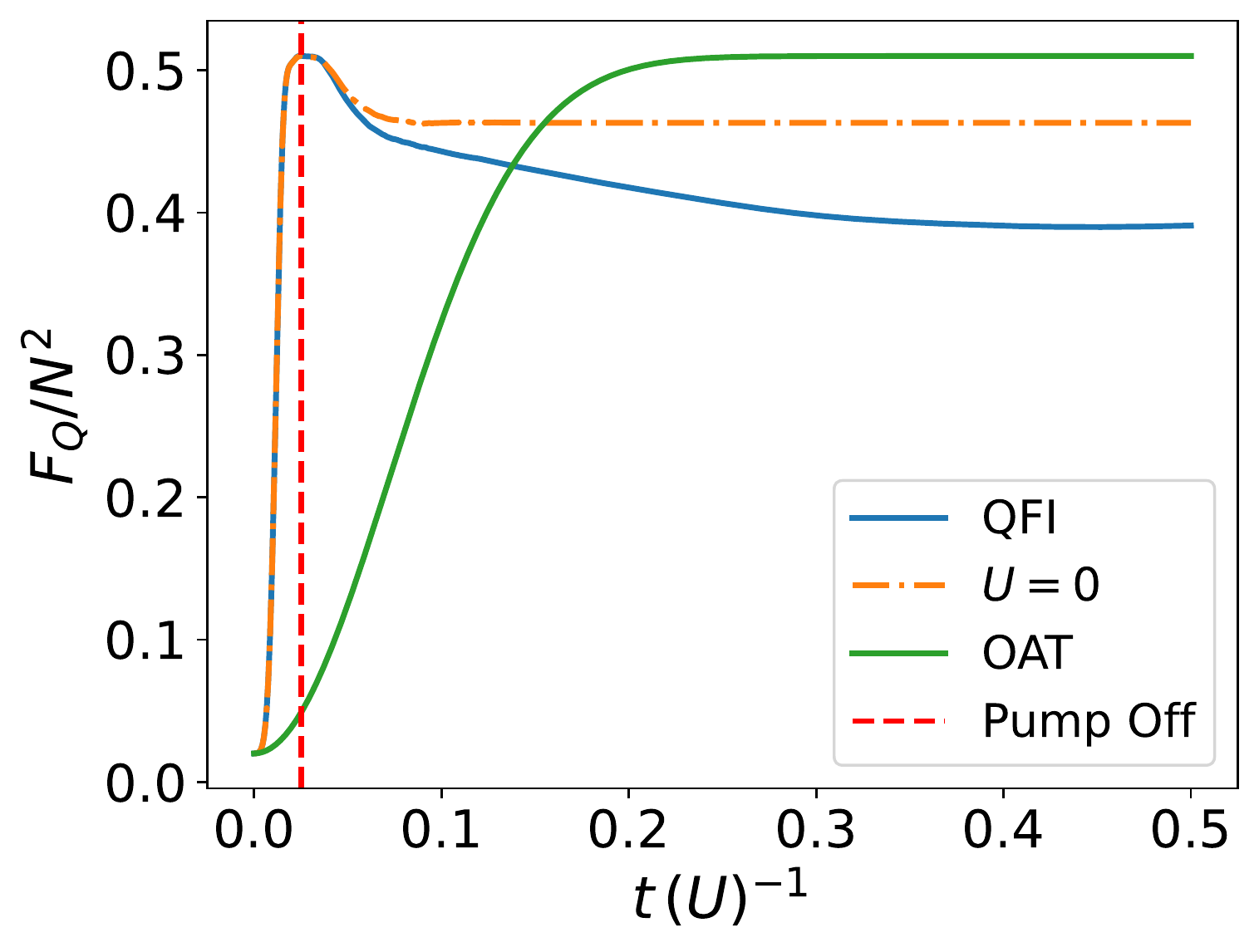}
     \includegraphics[width=0.49\linewidth]{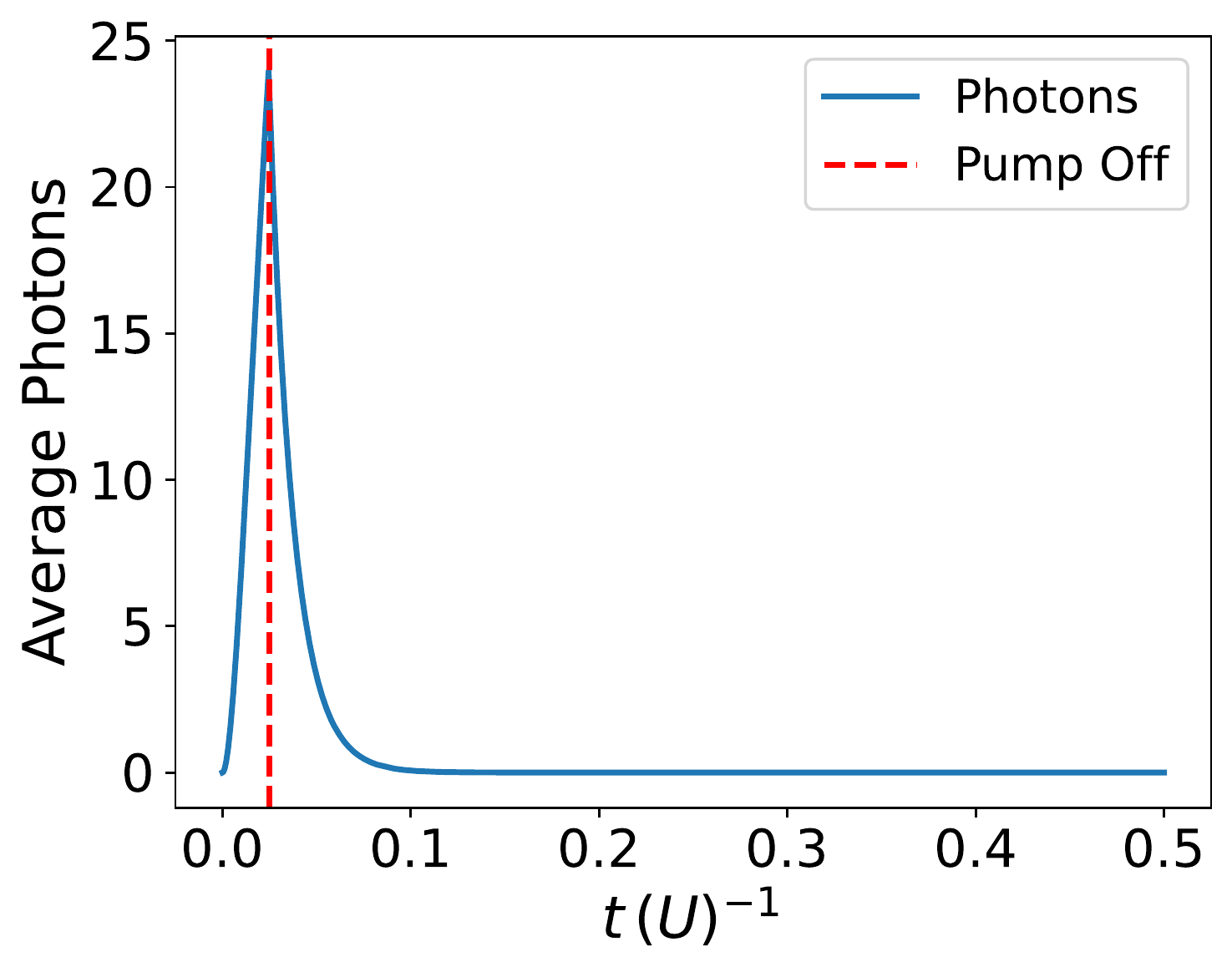}
     \caption{\textbf{Left}: Time evolution of $F_Q$ with finite duration of the pump for the loss $\kappa \simeq 78$ such that $\kappa\tau_Q=1$. The solid blue curve shows the evolution of $F_Q$, normalized by $N^2$. The vertical red dashed line indicates the moment when the pump $\eta$ is abruptly switched off. The initial value of pump $\eta$ is 320. The orange dot-dashed line represents the QFI dynamics when the on-site interaction energy $U=0$. The green curve corresponds to the one-axis twisting (OAT) model, serving as a benchmark for entanglement growth. 
     %The results highlight the persistence of significant entanglement within the atomic ensemble even after the pump is switched off and when losses $\kappa \tau_Q \sim 1$. 
     \textbf{Right}: Time evolution of the cavity photon number $\av{\hat{a}^\dag \hat{a}}$, showing rapid depletion following the cessation of the pump. The system parameters are: $U=W_0=\Delta_c=1$, $N=50$ atoms.}
     \label{fig:QFI_pump_dynamics}
\end{figure}

During the dynamics, we monitor the QFI and identify its maximum value. At the point of maximum $F_Q$, the external pump is abruptly switched off to investigate the persistence of entanglement in the atomic subsystem. Remarkably, our results demonstrate that $F_Q$ remains significantly large over longer timescales, even as the cavity photons are rapidly depleted. This sustained high QFI signifies robust many-body entanglement within the atomic ensemble, resilient to the loss of photons. We compare our results to the simulations with $U=0$ case (orange dot-dashed line) in order to show that finite atomic interactions are not responsible for the persistance of high $F_Q$

These findings suggest that the entangled atomic states retain their enhanced metrological properties despite the transient nature of the cavity field. The persistence of high QFI in the absence of continuous pumping suggests the potential of such states for quantum-enhanced sensing and precision measurements. The accompanying simulations, presented in Fig.~\ref{fig:QFI_pump_dynamics}, supports the stability of entanglement and its utility for metrological applications even with photon losses.

%\twtodo{@Sankalp: Can you upload the figure into SM with pump switched off at finite $t \sim \tau_Q$ which you showed in the email.}

\section{Optimal Measurement}

\begin{figure}[t!]
    \centering
    \includegraphics[width=0.31\linewidth]{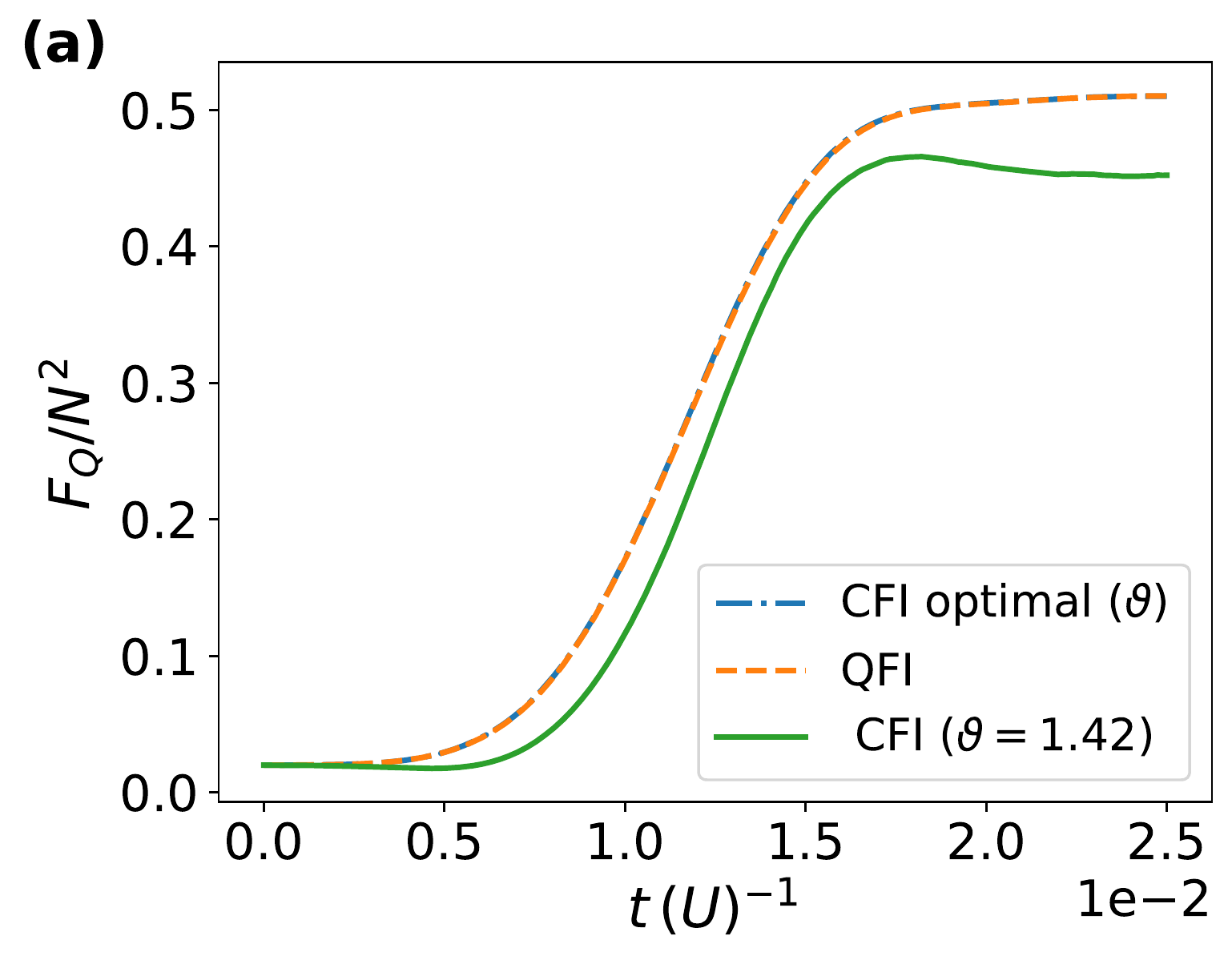}
    \includegraphics[width=0.31\linewidth]{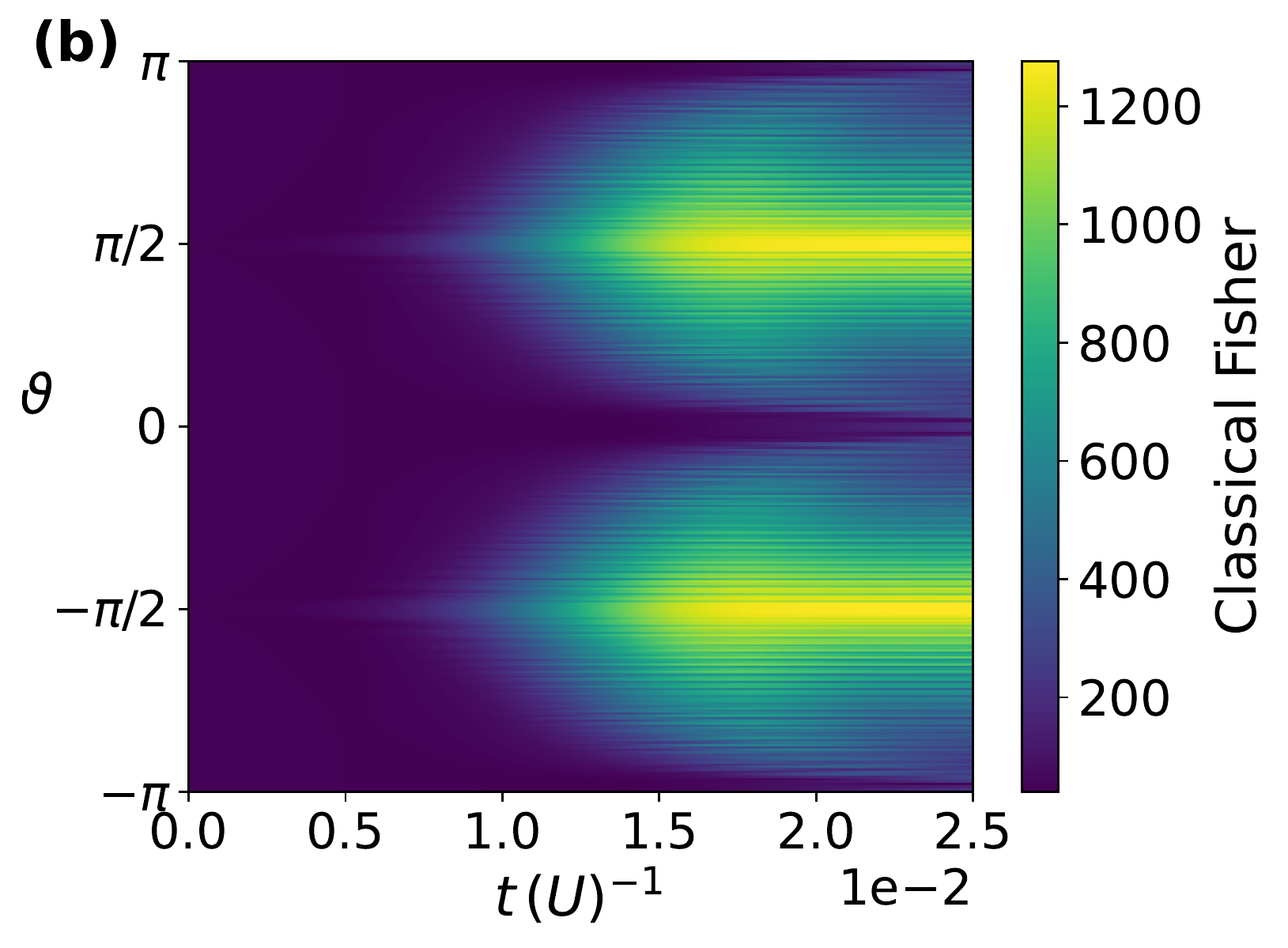}
    \includegraphics[width=0.31\linewidth]{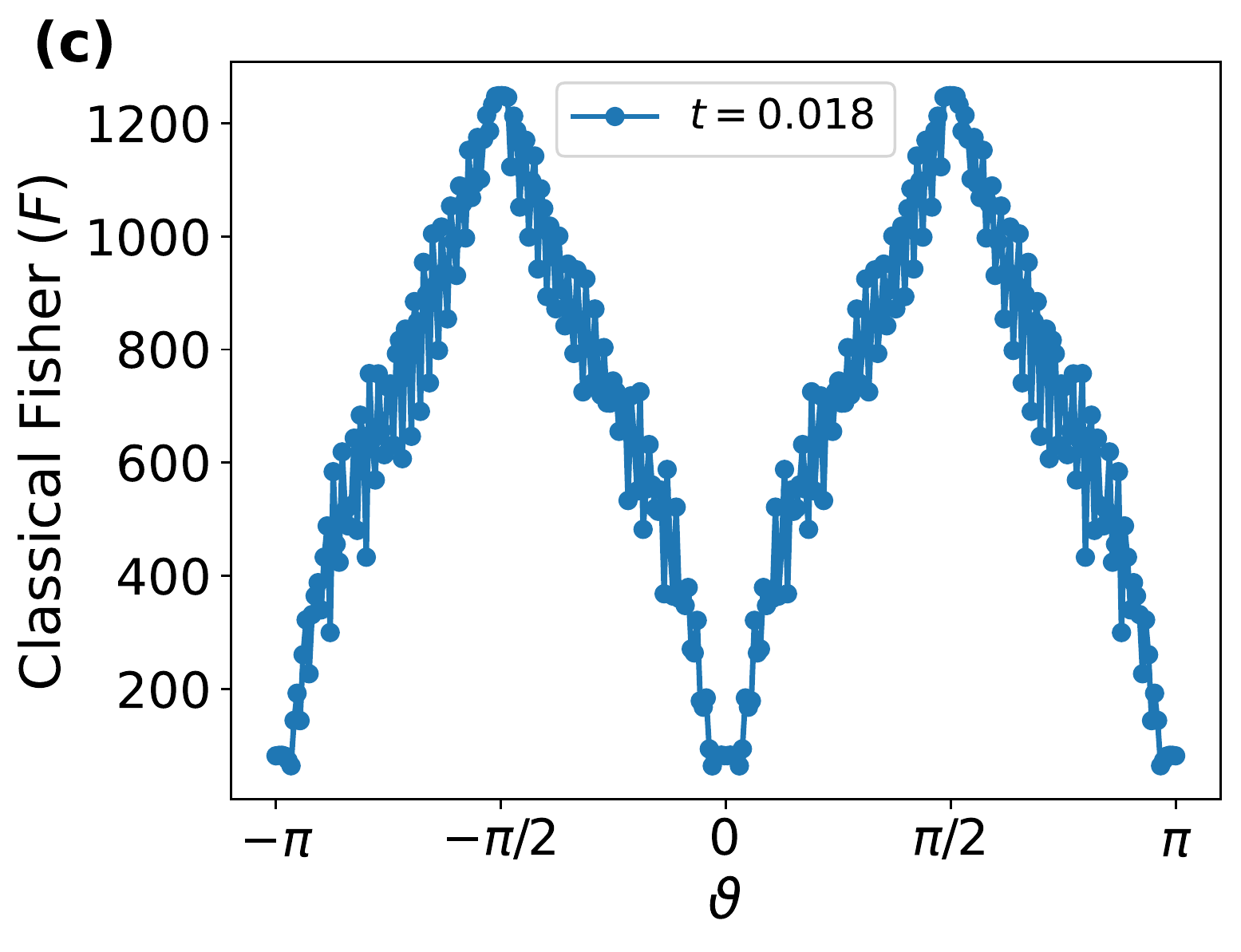}
    \caption{\textbf{(a):} Comparison of classical Fisher information  $F$ and Quantum Fisher Information $F_Q$ as a function of time. The dynamics presented here are for the pumped case. The fixed rotation angle $\vartheta \approx \frac{\pi}{2} \approx 1.56$ ensures that $F \approx F_Q$, even under losses  $\kappa \simeq 78$ such that $\kappa \tau_Q \sim 1$. \textbf{(b):} Dynamics of $F$ as a function of the rotation angle $\vartheta$ and time $t$. \textbf{(c):}  Variation of classical Fisher Information $F$ with rotation angle $\vartheta$ at a fixed time $t = 0.018$. Peaks in the Fisher Information indicate angles where the sensitivity to parameter estimation is maximized, highlighting optimal values of $\vartheta$ for precise measurements at this specific time. The system parameters are: $U=W_0=\Delta_c=1$, $N=50$ atoms and pump $\eta= 320$. 
    }
    \label{fig.optimal_measurement}
\end{figure}

In experimental settings, measurements yield classical Fisher information (CFI) $F$, which reflects the information accessible through specific measurement strategies. Below, we demonstrate that by selecting a fixed rotation angle $ \vartheta \approx \frac{\pi}{2} $ for all times, the classical Fisher information $F$ recovers the same value as the Quantum Fisher Information $F_Q$, even in the presence of losses, $\kappa \tau_Q \sim 1$.
This equivalence can be seen clearly in the attached figure (Fig.~\ref{fig.optimal_measurement}a), where $F$ is indistinguishable from $F_Q$ over time.  For comparison we also take a non-optimal value of the angle $\vartheta$ slightly shifted from $\pi/2$ by about $8\%$. In Fig.~\ref{fig.optimal_measurement}b we plot the CFI as a function of the variables $\vartheta$ and time $t$, showing that the maximum values in the CFI vs $\vartheta$ remain fixed in time. Finally, in Fig.~\ref{fig.optimal_measurement}c we plot the cut along $\vartheta$ for the fixed time $t$, demonstrating that for most of the angles $\vartheta$ apart from $\vartheta\approx\pm\pi$ and  $\vartheta\approx 0$ the values of CFI are larger than the shot-noise limit, certifying measurable entanglement.  

Our findings regarding the optimal measurement are particularly useful because it allows experimentalists to extract the full metrological advantage of the entangled atomic state using a simple and fixed measurement scheme. By avoiding the need for complex or time-dependent measurement protocols, this approach simplifies experimental implementation and enhances the feasibility of using many-body entangled states for quantum metrology applications.

\section{Finite tunneling rate $J\neq0$}

\begin{figure}[t!]
     \centering
     \includegraphics[width=0.49\linewidth]{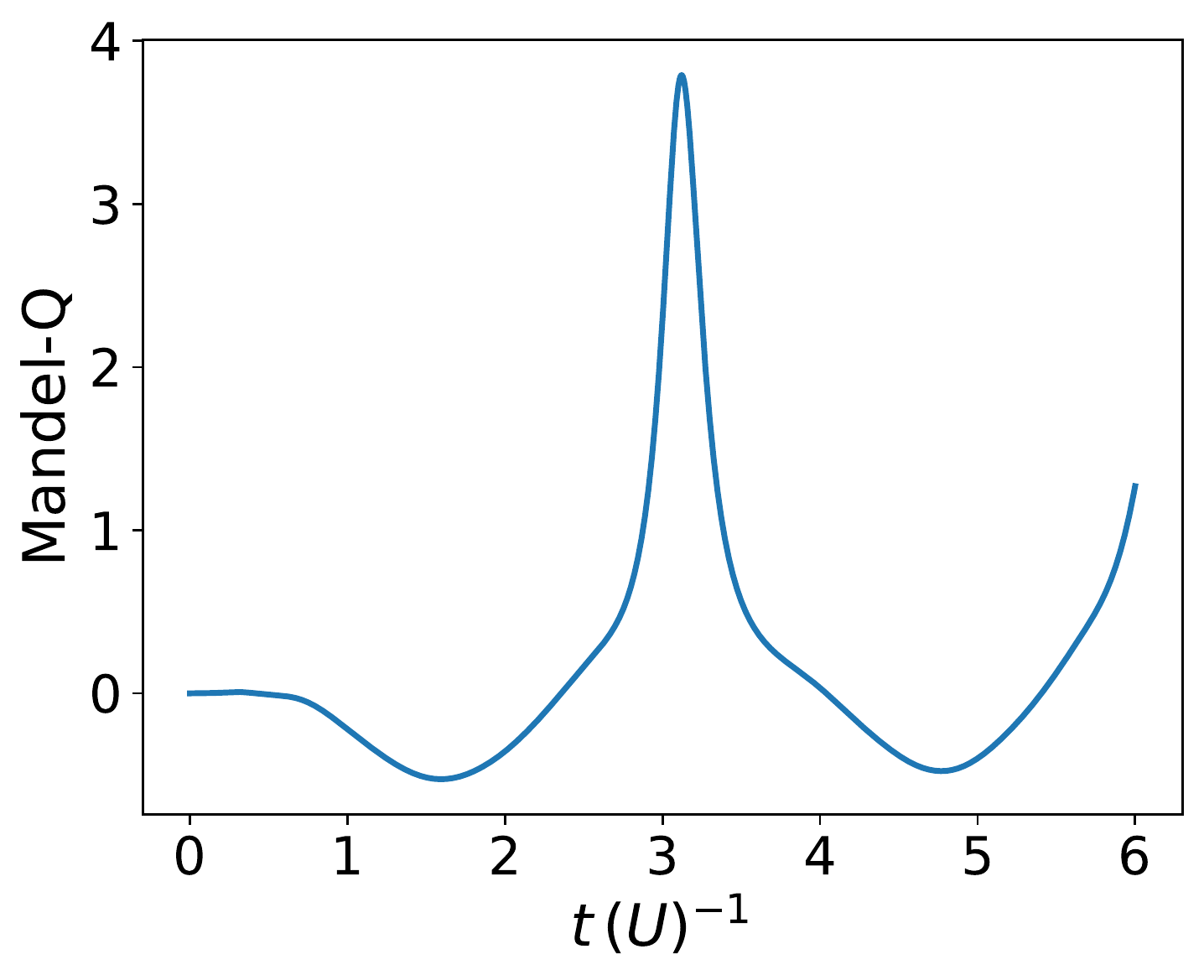}
     \includegraphics[width=0.49\linewidth]{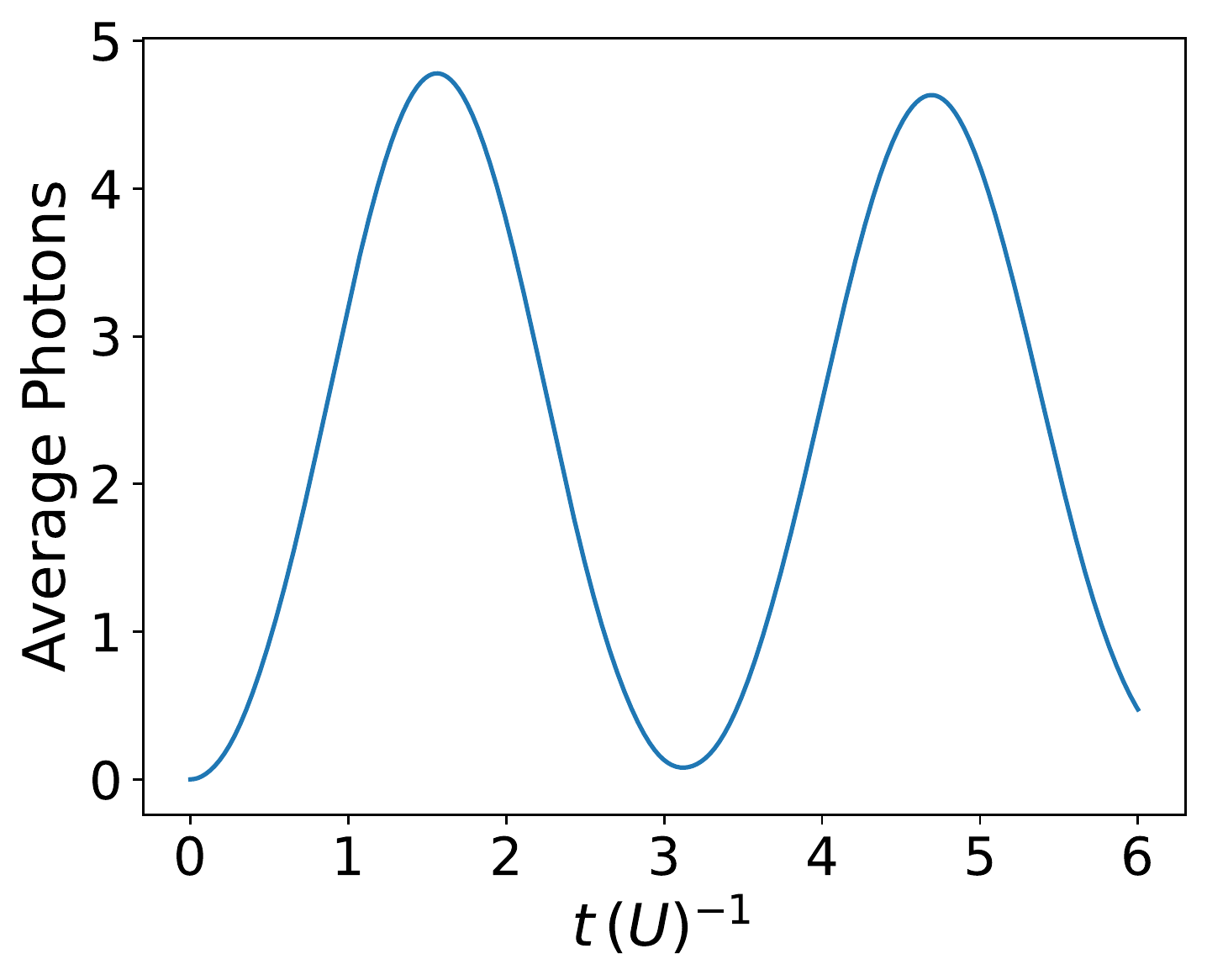}
     \caption{\textbf{Left}: Time evolution of the Mandel  $Q$ parameter. Here we start with a vacuum state of photons populated by a fixed pump $(\eta)$. The plot shows a transition between classical and non-classical behavior of the photon field, with classical regions around peaks (positive $Q$) and dips indicating regions of enhanced non-classicality corresponding to photon number squeezing (negative $Q$). \textbf{Right}: Average number of photons for the considered time. We did not take into account the photonic losses in the system. The system parameters are: $(U, W_0, \Delta_c, J) = (0.2, \, 0.5, \, 3.25, \, 2)$, $N= 20 $ atoms and pump $\eta=2$. }
     \label{fig:Mandel Q}
\end{figure}

In the case considered in the main text, where tunneling is absent, tracing out atomic degrees of freedom leaves the photonic state in the form of a classical mixture of coherent states. Therefore, without tunneling, the photons behave classically. For illustrative purposes we present Fig.~\ref{fig:Mandel Q}, where we have found numerically that with finite tunneling amplitude, the Mandel $Q$ parameter becomes negative, signaling non-classical states of light.
The Mandel $Q$  parameter is defined as: 
$Q = \frac{\langle (\Delta n)^2 \rangle - \langle n \rangle}{\langle n \rangle}
$ where $\langle (\Delta n)^2 \rangle = \langle n^2 \rangle - \langle n \rangle^2$  is the variance of the photon number~$n$.

\twocolumngrid

\end{document}